\pdfoutput=1
\documentclass[aps, prd, floats, floatfix, superscriptaddress, onecolumn,PRD,nofootinbib,amsmath,amssymb]{revtex4}

\usepackage[T1]{fontenc}
\usepackage[utf8]{inputenc}
\usepackage{lmodern}
\usepackage{lipsum}
\usepackage{enumerate}

\usepackage{MnSymbol}
\usepackage{amssymb}
\usepackage{amsfonts}
\usepackage{eufrak}

\usepackage[dvipsnames, usenames]{xcolor}
\usepackage{graphicx}
\usepackage{xspace}
\usepackage[normalem]{ulem}
\usepackage{bm}
\usepackage{phoenician}
\usepackage{appendix}

\definecolor{linkcolor}{rgb}{0.0,0.3,0.5}
\usepackage[hypertexnames=false, unicode, colorlinks=true, linkcolor=linkcolor,
citecolor=linkcolor, filecolor=linkcolor,urlcolor=linkcolor,
pdfusetitle]{hyperref}
\usepackage[all]{hypcap}

\usepackage{microtype}

\usepackage[english]{babel}
\usepackage{blindtext}

\usepackage{tikz}
\usetikzlibrary{shapes,snakes}
\usetikzlibrary{arrows,intersections,patterns,decorations.markings,shapes.geometric}
\usetikzlibrary{calc,fadings,decorations.pathreplacing,positioning}
\usetikzlibrary{decorations.pathmorphing}
\tikzset{snake it/.style={decorate, decoration=snake}}

\usepackage{pgfplots}
\usepgfplotslibrary{colormaps}
\usetikzlibrary{intersections,patterns,pgfplots.fillbetween}

\tikzset{->-/.style={decoration={
  markings,
  mark=at position .5 with {\arrow{>}}},postaction={decorate}}
}
\tikzset{-<-/.style={decoration={
  markings,
  mark=at position .5 with {\arrow{<}}},postaction={decorate}}
}

\tikzset{%
  >=latex, 
  inner sep=0pt,%
  outer sep=2pt,%
  mark coordinate/.style={inner sep=0pt,outer sep=0pt,minimum size=3pt,
    fill=black,circle}%
}

\usepackage{tikz-3dplot}

\renewcommand{\vec}[1] {\bm{#1}}
\newcommand{\vhat}[1]{\vec{\hat{#1}}}

\newcommand{\fnl}{f_\text{NL}}
\newcommand{\Mpc}{{\rm Mpc}}
\newcommand{\pr}{^{\prime}}

\newcommand{\kf}{k_f}

\newcommand{\PA}{P^{\rm A}}
\newcommand{\PS}{P^{\rm S}}

\newcommand{\ks}{k_1}   
\newcommand{\vecks}{\vec{k}_1}
\newcommand{\kl}{k_3}   
\newcommand{\veckl}{\vec{k}_3}
\newcommand{\vhatkl}{\vhat{k}_3}

\newcommand{\Koverk}{\left(\frac{\kl}{\ks}\right)}
\newcommand{\mus}{\mu_1}
\newcommand{\mul}{\mu_3}
\newcommand{\musl}{\mu_{13}}

\newcommand{\tA}{A}
\newcommand{\tB}{B}

\newcommand{\bltA}{b_1^{\tA}}
\newcommand{\bltB}{b_1^{\tB}}
\newcommand{\bstA}{b_2^{\tA}}
\newcommand{\bstB}{b_2^{\tB}}
\newcommand{\bKtA}{b_{K^2}^{\tA}}
\newcommand{\bKtB}{b_{K^2}^{\tB}}
\newcommand{\bphitA}{b_{\phi}^{\tA}}
\newcommand{\bphitB}{b_{\phi}^{\tB}}
\newcommand{\bphideltatA}{b_{\phi\delta}^{\tA}}
\newcommand{\bphideltatB}{b_{\phi\delta}^{\tB}}
\newcommand{\bll}{\left(\bltA-\bltB\right)}
\newcommand{\bss}{\left(\bstA-\bstB\right)}
\newcommand{\bKK}{\left(\bKtA-\bKtB\right)}
\newcommand{\bls}{\left(\bltA\bstB-\bltB\bstA\right)}
\newcommand{\blK}{\left(\bltA\bKtB-\bltB\bKtA\right)}
\newcommand{\dPk}{\frac{\partial P(\ks)}{\partial \ks}}

\newcommand{\dlogPkdlogk}{\frac{\partial\log P(\ks)}{\partial\log\ks}}
\newcommand{\dPsecond}{\frac{\ks^2}{P(\ks)}\frac{\partial^2P(\ks)}{\partial\ks^2}}
\newcommand{\dPthird}{\frac{\ks^3}{P(\ks)}\frac{\partial^3P(\ks)}{\partial\ks^3}}

\newcommand{\dTfirst}{\frac{\partial\log T(\ks)}{\partial\log\ks}}

\newcommand{\blphi}{\left(\bltA\bphitB-\bltB\bphitA\right)}
\newcommand{\bphiK}{\left(\bphitA\bKtB-\bphitB\bKtA\right)}
\newcommand{\bphis}{\left(\bphitA\bstB-\bphitB\bstA\right)}
\newcommand{\blphidelta}{\left(\bltA\bphideltatB-\bltB\bphideltatA\right)}
\newcommand{\bphi}{\left(\bphitA-\bphitB\right)}
\newcommand{\bphidelta}{\left(\bphideltatA-\bphideltatB\right)}

\newcommand{\Qfenicio}{\text{\phncfamily\textphnc{q}}}

\newcommand{\biascombination}{\beth} 

\definecolor{darkred}{RGB}{175,0,0}
\definecolor{darkblue}{RGB}{14,0,185}

\begin{document}

\title{
Antisymmetric galaxy cross-correlations in and beyond $\Lambda$CDM
}

\newcommand\alvisehome{
\affiliation{Dipartimento di Fisica Galileo Galilei, Universit\` a di Padova, I-35131 Padova, Italy}
\affiliation{INFN Sezione di Padova, I-35131 Padova, Italy}
\affiliation{INAF-Osservatorio Astronomico di Padova, Italy}
\affiliation{Theoretical Physics Department, CERN, 1 Esplanade des Particules, 1211 Geneva 23,
Switzerland}}

\newcommand\elehome{
\affiliation{Dipartimento di Fisica Galileo Galilei, Universit\` a di Padova, I-35131 Padova, Italy}
\affiliation{INFN Sezione di Padova, I-35131 Padova, Italy}
}

\newcommand\nichome{
\affiliation{Dipartimento di Fisica Galileo Galilei, Universit\` a di Padova, I-35131 Padova, Italy}
\affiliation{INFN Sezione di Padova, I-35131 Padova, Italy}
\affiliation{INAF-Osservatorio Astronomico di Padova, Italy}
}

\author{Eleonora Vanzan}
\email{eleonora.vanzan@phd.unipd.it}
\elehome

\author{Alvise Raccanelli}
\email{alvise.raccanelli.1@unipd.it}
\alvisehome

\author{Nicola Bartolo}
\email{nicola.bartolo@pd.infn.it}
\nichome

\begin{abstract}

Many different techniques to analyze galaxy clustering data and obtain cosmological constraints have been proposed, tested and used. Given the large amount of data that will be available soon, it is worth investigating new observables and ways to extract information from such datasets.
In this paper, we focus on antisymmetric correlations, that arise in the cross-correlation of different galaxy populations when the small-scale power spectrum is modulated by a long-wavelength field. In $\Lambda$CDM this happens because of nonlinear clustering of sources that trace the underlying matter distribution in different ways. Beyond the standard model, this observable is sourced naturally in various new physics scenarios.

We derive, for the first time, its complete expression up to second order in redshift space, and show that this improves detectability compared to previous evaluations at first order in real space. Moreover, we explore a few potential applications to use this observable to detect models with vector modes, or where different types of sources respond in different ways to the underlying modulating long mode, and anisotropic models with privileged directions in the sky.
This shows how antisymmetric correlations can be a useful tool for testing exotic cosmological models.

\end{abstract}

\maketitle

\section{Introduction}
\label{sec:intro}

Current and future galaxy surveys are expected to map the Large-Scale Structure (LSS) of the Universe with unprecedented detail, providing us with catalogs of different tracers of the underlying dark matter field. Observing different types of sources will allow for the use of the multi-tracer approach, which promises to be a winning strategy to beat down cosmic variance~\cite{Seljak2008,McDonald2008,White2008,Abramo2013,Viljoen2021,Abramo2021}. 

Recently, there have been many efforts to improve the available statistical tools for analyzing the LSS, through a more accurate modeling of e.g.,~observational, small-scale and general-relativistic corrections. However, it is also worth investigating completely new avenues, which may both complement existing observables and be better-suited to test specific cosmological models. A recently developed observable~\cite{Jeong2012}, searches for imprints on the two-point statistics from primordial fossil fields. These fields could be either scalar, vector, or tensor modes, and they would induce local departures from an otherwise statistically isotropic two-point function.

An extension to this observable has been proposed in~\cite{Dai2015}, to exploit the benefits of the multitracer technique: the antisymmetric part of the galaxy cross-correlation, which will be non-vanishing in the presence of two different tracers. If one considers two galaxies drawn from the same population, separated by a distance $\vec{r}$, the two-point auto-correlation function is symmetric under the inversion $\vec{r} \mapsto -\vec{r}$. But if the two galaxies belong to different populations, with different biasing and evolution properties, then their cross-correlation function may not be symmetric under this exchange.
Such an antisymmetric term is generated in standard $\Lambda$CDM, as shown in~\cite{Dai2015}, because of biased nonlinear clustering: it arises from the fact that the two populations trace the dark matter field in different ways, i.e.,~they have different bias parameters.

An antisymmetric contribution can also arise from exotic new physics, e.g.,~the presence of vector fields that leave an imprint on the galaxy clustering.
This work, starting from the idea sketched in~\cite{Dai2015}, derives a more complete expression for the antisymmetric part of galaxy correlations, including redshift-space distortions, a more detailed modeling for the galaxy bias, and the effect of primordial non-Gaussianity.
Then, for the first time, it investigates the detectability of such an observable by future galaxy surveys.
Furthermore, it explores a few possible exotic physics models that could be tested using this new observable.

\section{Antisymmetric galaxy correlation}
\label{sec:anti}

The two-point correlation function is one of the most widely used summary statistics in large-scale structure surveys.
In this context, it is convenient to work in Fourier space (see e.g.,~\cite{Peebles1980}), where modes are independent and the covariance matrix is diagonal\footnote{Assuming certain approximations, namely the plane-parallel and equal time; for a detailed analysis of this, see recent discussions in~\cite{Raccanelli2023a,Raccanelli2023b,Gao2023}.}.
Alternative approaches are, e.g.,~the two-point function in configuration space (see e.g,~\cite{Szalay1997,Matsubara1999,SDSS2005,Bertacca2012}), spherical-Fourier Bessel~\cite{Heavens1997,2dFGRS2004,Yoo2013,Gebhardt2021}, the angular power spectrum~\cite{Yu1969,Peebles1973,Fisher1993,Yoo2008,Yoo2009,Bonvin2011,Challinor2011}, the frequency-angular power spectrum~\cite{Raccanelli2023a,Raccanelli2023b,Gao2023}.

The two-point auto-correlation function is usually assumed to inherit the properties of statistical homogeneity and isotropy of the FLRW Universe. However, it was pointed out in~\cite{Jeong2012} that in principle the two-point function may depend on the orientation of the two points being correlated and/or on their position in space. Such a signature can be decomposed into scalar, vector, and tensor components, and it can be parameterized as
\begin{equation}
\label{eq:one}
    \langle \delta(\vec{k}_1)\delta(\vec{k}_2) \rangle_h = (2\pi)^3 \delta^{(3)}(\vec{k}_1+\vec{k}_2) P(k_1) +\int\frac{d^3\veckl}{(2\pi)^3} \sum_p f_p(k_1,k_2,\vec{k}_1\cdot\vec{k}_2) h^*_p(\veckl) \epsilon_{ij}^p(\veckl) k_1^i k_2^j (2\pi)^3 \delta^{(3)}(\vec{k}_1+\vec{k}_2+\veckl) \, .
\end{equation}
The second term above is a correlation induced by a perturbation $h$ with polarization $p$ and wavevector $\veckl$, where $\veckl$ is a long-wavelength mode that is modulating the two-point function and $\vec{k}_1$, $\vec{k}_2$ are two short-wavelength modes. The sum on $p$ runs over the six possible basis tensors for a symmetric tensor. Intuitively, this represents a power spectrum sitting on top of a long-wavelength mode, and it is related to the squeezed bispectrum as $\lim_{\kl \ll \ks} B_p(\vec{k}_1,\vec{k}_2,\veckl) = P_p(\kl) f_p(k_1,k_2,\vec{k}_1\cdot\vec{k}_2) \epsilon^p_{ij}(\veckl) k_1^i k_2^j$~\cite{Simonovic2014}. In~\cite{Dimastrogiovanni2014, Dimastrogiovanni2015} this formalism has been applied to study the imprint of primordial fossil fields from inflation on the large-scale structure.

In~\cite{Dai2015} this parametrization has been generalized to the case of the two-point cross-correlation function, by including the three additional degrees of freedom that are related to the antisymmetric part and in principle arise for different tracers
\begin{equation}
    \label{eq:two}
    \langle \delta_{\tA}(\vec{k}_1)\delta_{\tB}(\vec{k}_2) \rangle_h = \int\frac{d^3\veckl}{(2\pi)^3} \sum_p f_p(k_1,k_2,\vec{k}_1\cdot\vec{k}_2) h^*_p(\veckl) \hat{\epsilon}_p \cdot (\vec{k}_1-\vec{k}_2) (2\pi)^3 \delta^{(3)}(\vec{k}_1+\vec{k}_2+\veckl) \, ,
\end{equation}
where the sum on $p$ runs over the three polarizations $p=L,x,y$ that is, a longitudinal mode and two vector modes. Choosing $\hat{\epsilon}_L(\veckl)=\vhatkl$, then $\hat{\epsilon}_{x,y}(\veckl)$ are two other unit vectors, orthogonal to $\vhatkl$ and to each other.

No assumption has been made so far on the nature of the long mode that modulates the power spectrum. It could be generated by new physics, and some of these exotic scenarios will be explored in Section~\ref{sec:exoticmodels}. But an antisymmetric contribution to the two-point cross-correlation arises even in pure $\Lambda$CDM, due to the nonlinear clustering of biased tracers.

In order to appreciate the underlying physics, it is useful to briefly recall the framework studied in~\cite{Dai2015}. The abundance of tracers is in general a nonlinear function of the local mass density: for simplicity, let $\delta_X(\vec{x}) = b_X \delta(\vec{x}) +c_X \left[\delta(\vec{x})\right]^2 +\dots$ with $b_X$ the linear bias parameter and $c_X$ the nonlinear bias parameter~\cite{Desjacques2016}. Such a nonlinear relation implies a nonvanishing three-point function
\begin{equation}
    \langle \delta_{\tA}(\vec{k}_1)\delta_{\tB}(\vec{k}_2)\delta(\veckl) \rangle = 2P(\kl) \left[ b_{\tA}c_{\tB}P(k_1) +b_{\tB}c_{\tA}P(k_2) \right] (2\pi)^3\delta^{(3)}(\vec{k}_1+\vec{k}_2+\veckl) \, .
\end{equation}
After antisymmetrization in $\vec{k}_1 \longleftrightarrow \vec{k}_2$, and after taking the squeezed limit $k_1,k_2 \gg \kl$
\begin{equation}
    \frac{ B^{\rm A}(\vecks,\veckl) }{ P(\kl) } = \left(b_{\tB}c_{\tA} -b_{\tA}c_{\tB}\right) \frac{\partial P(\ks)}{\partial \ks} \frac{\vecks\cdot\veckl}{\ks} \, .
\end{equation}
This sources the antisymmetric part of the modulation of the two-point function due to nonlinear biased clustering, as
\begin{equation}
    \frac{ \langle\delta_{\tA}(\vec{k}_1)\delta_{\tB}(\vec{k}_2)\rangle -\langle\delta_{\tB}(\vec{k}_1)\delta_{\tA}(\vec{k}_2)\rangle }{2} = \int\frac{d^3\veckl}{(2\pi)^3} \left(\frac{B^{\rm A}(\vecks,\veckl)}{P(\kl)}\right) \delta^*(\veckl) (2\pi)^3 \delta^{(3)}(\vec{k}_1+\vec{k}_2+\veckl) \, .
\end{equation}
Comparing to the general parametrization of the two-point function,~\eqref{eq:two} allows to recognize the presence of a longitudinal mode $p=L$, with
\begin{equation}
\label{eq:fLMK}
    f_L = \frac{1}{2} \left(b_{\tB}c_{\tA} -b_{\tA}c_{\tB}\right) \frac{\partial P(\ks)}{\partial \ks} \frac{\kl}{\ks} \, .
\end{equation}
Physically, this signal describes the small-scale clustering of tracers in the presence of a low-pass-filtered density field. In other words, it is related to the power spectrum of the two tracers on top of a long underlying mode, $\kl$.

To estimate the impact of this signal compared to the symmetric one, consider the ratio of the antisymmetric part
\begin{equation}
    \PA = \left(b_{\tB}c_{\tA}-b_{\tA}c_{\tB}\right) \frac{\partial P(\ks)}{\partial \ks} \delta^*(\veckl) \frac{\vecks\cdot\veckl}{\ks} \, ,
\end{equation}
to the symmetric part in the squeezed configuration
\begin{equation}
    \PS = \left(b_{\tB}c_{\tA}+b_{\tA}c_{\tB}\right) \delta^*(\veckl) \left[P(k_1)+P(k_2)\right] \simeq \left(b_{\tB}c_{\tA}+b_{\tA}c_{\tB}\right) 2 \delta^*(\veckl) P(\ks) \left(1+\frac{1}{2P(\ks)}\frac{\partial P(\ks)}{\partial \ks}\frac{\vecks\cdot\veckl}{\ks}\right) \, ,
\end{equation}
that gives
\begin{equation}
    \frac{\PA}{\PS} = \cfrac{ \left(b_{\tB}c_{\tA}-b_{\tA}c_{\tB}\right) }{ \left(b_{\tB}c_{\tA}+b_{\tA}c_{\tB}\right)} \frac{1}{2P(\ks)} \frac{\partial P(\ks)}{\partial \ks} \frac{\vecks\cdot\veckl}{\ks} +\mathcal{O}(\kl^2) \, .
\end{equation}
Since $\ks$ are small-scale modes, $P(\ks)$ can be taken to be a powerlaw $P(\ks) \propto \ks^n$ with $n \simeq -3$. Then, an approximate estimate will be
\begin{equation}
    \frac{\PA}{\PS} = \frac{\left(b_{\tB}c_{\tA}-b_{\tA}c_{\tB}\right)}{\left(b_{\tB}c_{\tA}+b_{\tA}c_{\tB}\right)} \frac{n}{2} \frac{\kl}{\ks} \, .
\end{equation}
The ratio is therefore suppressed as $\kl/\ks$; however, it could be that, for a particular combination of the bias parameters, the bias-dependent pre-factor boosts the signal.

An improved, more detailed modeling of the signal must include the full second order kernels, containing the physics of gravitational evolution and nonlinear clustering, as well as a more accurate modeling of the bias, with the full basis of bias operators beyond the local-in-matter density expansion up to second order~\cite{Desjacques2016}. The rest of the paper derives such an expression and argues that not only will it be more correct, but also improve its detectability.

\subsection{Redshift-space distortions}
\label{sec:RSD}

Observations do not happen in real space, but in the so-called {\it observed} space, where one has to account for all the lightcone and perturbation effects; this is commonly called redshift-space when accounting for the (generally) dominant redshift-space distortions (RSD) caused by peculiar velocities.
In this case, the clustering pattern of objects is modified by peculiar velocities~\cite{Kaiser1987, Hamilton1997}. Since they are sensitive to the line of sight component of peculiar velocities, RSD are a radial effect, and for the case of interest, there is no transverse vector perturbation that can feed the $p=x,y$ polarizations. Therefore, the expression for the antisymmetric cross-correlation will still contain the $L$ polarization only, but it will be enriched by additional contributions.

Up to second order in fluctuations, one can write the density contrast of the $X$-th tracer as
\begin{equation}
    \delta_X(\vec{k}) = Z_1^{X}(\vec{k}) \delta(\vec{k}) +\int\frac{d^3\vec{p}d^3\vec{q}}{(2\pi)^3} \delta^{(3)}(\vec{k}-\vec{p}-\vec{q}) Z_2^{X}(\vec{p},\vec{q}) \delta(\vec{p}) \delta(\vec{q}) \, ,
\end{equation}
where $Z_{1,2}$ are the first and second order kernels.

The observable of interest here will require the calculation of both the 2- (power spectrum) and 3- point (bispectrum) statistics.
The bispectrum is
\begin{equation}
\begin{split}
    \langle \delta_{s,\tA}(\vec{k}_1) \delta_{s,\tB}(\vec{k}_2) \delta_s(\veckl) \rangle = 2 \Big\{ &P(k_1)P(k_2) Z_1^{\tA}(\vec{k}_1) Z_1^{\tB}(\vec{k}_2) Z_2(-\vec{k}_1,-\vec{k}_2) \\
    + &P(k_2)P(\kl) Z_2^{\tA}(-\vec{k}_2,-\veckl) Z_1^{\tB}(\vec{k}_2) Z_1(\veckl) \\
    + &P(k_1)P(\kl) Z_1^{\tA}(\vec{k}_1) Z_2^{\tB}(-\vec{k}_1,-\veckl)  Z_1(\veckl) \Big\} (2\pi)^3\delta^{(3)}(\vec{k}_1+\vec{k}_2+\veckl) \, .
\end{split}
\end{equation}

In order to calculate the expression for the RSD operator, connecting real- to resdhift- space observables, one has to compute the Jacobian of the transformation. The calculation can be found in literature, see e.g.~\cite{Hamilton1997,Szalay1997,Raccanelli2016}, and it reads
\begin{equation}
1+\delta_s(\vec{s}) = \left[1+\delta(\vec{r})\right] \left(1+\frac{\partial v_r}{\partial r}\right)^{-1} \left(1+\frac{v_r}{r}\right)^{-2} \frac{\Bar{n}(\vec{r})}{\Bar{n}\left(\vec{r}+v_r(\vec{r})\vhat{r}\right)} \, .
\end{equation}
The relation between redshift space and real space is $\vec{s}=\vec{r}+v_r(\vec{r})\vhat{r}$, with $v_r=\vec{v}\cdot\vhat{r}/(aH)$, and $\Bar{n}$ is the average number density of tracers. Keeping all terms, one obtains, to second order
\begin{equation}
\begin{split}
\label{eq:RSD1}
    \delta_s \simeq& \, \delta -\frac{1}{\mathcal{H}} \partial_r v -\frac{\alpha v}{\mathcal{H} r} +\frac{1}{2\mathcal{H}^2} \partial_r^2 v^2 -\frac{1}{\mathcal{H}} \partial_r(\delta v) -\frac{\alpha \delta v}{\mathcal{H} r} +\frac{\left( \alpha(\alpha-2)-\gamma \right) v^2}{\mathcal{H}^2 r^2} +\frac{1}{\mathcal{H}^2} \partial_r\left(\frac{\alpha v^2}{r}\right) \, ,
\end{split}
\end{equation}
\begin{equation}
\label{eq:RSD2}
    \alpha \equiv \frac{r}{\Bar{n}}\frac{\partial\Bar{n}}{\partial r} +2 \, , \qquad\qquad \gamma \equiv \frac{r^2}{2\Bar{n}}\nabla^2\Bar{n} -3 \, .
\end{equation}
Neglecting the Doppler term $v/r$ and selection effects\footnote{For the results in this work the effects of the Doppler term are neglected, as it generally dominates only on large scales, while the antisymmetric correlation has most of its signal on small scales. A more detailed investigation of the impact of Doppler and other terms on the antisymmetric correlation is undergoing. The complete kernels expressions are reported in Appendix~\ref{app:extra}.}, the relation between redshift-space and real-space density perturbation becomes
\begin{equation}
    \delta_s \simeq \delta -\frac{1}{\mathcal{H}} \partial_r v +\frac{1}{2\mathcal{H}^2} \partial_r^2 v^2 -\frac{1}{\mathcal{H}} \partial_r (\delta v) \, .
\end{equation}

The objects being correlated are biased tracers of the underlying dark matter distribution. The relation between the two can be described through the bias expansion~\cite{Desjacques2018}
\begin{equation}
    \delta_g = b_1\delta +\frac{b_2}{2}\delta^2 +b_{K^2}K^2 \, , \qquad\qquad K_{ij} = \left( \frac{\partial_i\partial_j}{\nabla^2}-\frac{1}{3}\delta_{ij} \right)\delta \, ,
\end{equation}

The second-order gravitational evolutions kernels $F_2$ and $G_2$ in an Einstein-de Sitter cosmology are
\begin{align}
    & F_2(\vec{p},\vec{q}) = \frac{5}{7} +\frac{2}{7} \frac{\left(\vec{p}\cdot\vec{q}\right)^2}{p^2q^2} +\frac{\vec{p}\cdot\vec{q}}{2pq} \left( \frac{p}{q}+\frac{q}{p} \right) \, , \\
    & G_2(\vec{p},\vec{q}) = \frac{3}{7} +\frac{4}{7} \frac{\left(\vec{p}\cdot\vec{q}\right)^2}{p^2q^2} +\frac{\vec{p}\cdot\vec{q}}{2pq} \left( \frac{p}{q}+\frac{q}{p} \right) \, .
\end{align}

The first and second order redshift space kernels for biased tracers are, in Fourier space
\begin{equation}
    Z_1(\vec{k}) = b_1 +f\mu^2 \, ,
\label{eq:Z1}
\end{equation}
\begin{equation}
\begin{split}
    Z_2(\vec{p},\vec{q}) =& \frac{b_2}{2} +b_1F_2(\vec{p},\vec{q}) +b_{K^2}\left(\mu_{pq}^2-\frac{1}{3}\right) +f\mu^2 G_2(\vec{p},\vec{q}) +\frac{k\mu f}{2}\left(\frac{\mu_p}{p}\left(b_1+f\mu_q^2\right) +\frac{\mu_q}{q}\left(b_1+f\mu_p^2\right)\right) \, .
\end{split}
\label{eq:Z2}
\end{equation}
The Dirac delta in the convolution enforces that $\vec{k}=\vec{p}+\vec{q}$. The cosines of the angles between wavevectors and the line of sight are $\mu=\vec{k}\cdot\vhat{n}/k$, and the cosine of the angle between the two wavevectors is $\mu_{pq}=\vec{p}\cdot\vec{q}/(pq)$.

Following the same procedure as above, the antisymmetric part of the cross-power spectrum will be sourced by the antisymmetrized bispectrum in the squeezed regime -- because the $n$-point function gets modulated by the $(n+1)$-point function with a soft momentum as in Equation~\eqref{eq:one},
\begin{equation}
\begin{split}
\label{eq:BA}
    \frac{ B^{\rm A}(\vecks,\veckl) }{ (1+f\mul^2)P(\kl) } =& \frac{1}{42} \Koverk \\
    & \Bigg\{ 3\bll f\mus \Big[ \ks\dPk \musl \left(7f\mus^3-14\musl\mul+\mus (4+10\musl^2-7f\mul^2)\right) \\
    & \qquad +2P(\ks) \left(-14 f\mus^3\musl+7f\mus^2\mul+3(3+4\musl^2)\mul+\mus\musl(-13-8\musl^2+7f\mul^2)\right) \Big] \\
    & +21\bss f\mus \left( \ks\dPk \mus\musl +2P(\ks) (-\mus\musl+\mul) \right) \\
    & +14\bKK f\mus \left(\ks\dPk \mus\musl (3\musl^2-1) -2P(\ks) (-4 \mus\musl+6\mus\musl^3+\mul-3\musl^2\mul)\right) \\
    & -21 \bls \ks\dPk \musl \\
    & +14 \blK \musl \left(6P(\ks)(\musl^2-1) +\ks\dPk (1-3\musl^2)\right) \Bigg\} \, .
\end{split}
\end{equation}

This is related to the antisymmetric component of the two point by $B^{\rm A}(\vec{k}_1,\vec{k}_2,\veckl) = P(\veckl) f_p^{\rm A}(\vec{k}_1,\veckl) \Hat{\epsilon}_p \cdot \left(\vec{k}_1-\vec{k}_2\right)$, with the long mode power spectrum $P(\veckl) = (1+f\mul^2)P(\kl)$.

Comparing with the general parametrization of Equation~\eqref{eq:two}, only the longitudinal vector mode is present $\Hat{\epsilon}_L(\veckl)=\vhatkl$, as expected. Indeed, as RSD are a radial effect, there is no physics that can excite the transverse vector modes.

Notice that, when including RSD, the signal is nonvanishing even in the case of purely linear bias. In particular, keeping RSD and linear bias only, the surviving signal from Equation~\eqref{eq:BA} would be
\begin{equation}
\label{eq:BAlinearbias}
    \frac{1}{2}\Koverk \left(\bltA-\bltB\right) P(\ks) f^2 \mus^2 \left[ 2 (-2\mus^2\musl+\mus\mul+\musl\mul^2) + \musl(\mus^2-\mul^2) \dlogPkdlogk \right] \, .
\end{equation}
This surviving contribution can be read as follows. The terms from RSD that are only sensitive to gravity cancel out because of equivalence principle, but mixed contributions that also contain the bias parameters do survive in the multitracer approach.

The antisymmetric cross-correlation is sensitive to the separation of scales between long and short wavelength modes; this is transparent in Equation~\eqref{eq:fLMK}, where the signal is directly proportional to the squeezing factor $\kl/\ks$. In the simple case of not modeling higher order effects that enters when the squeezing factor is reduced, this is unavoidably an arbitrary choice, that depends on the details of the analysis and on the specifications of the survey. An attempt to reduce this issue and model the full signal including the case of squeezing factor reduction is currently under development.
For now, as a benchmark, the rest of the work uses a squeezing factor of at least 10 between long and short modes; more details on this issue, and how the signal depends on it, will be addressed later.

\subsection{Primordial non-Gaussianity}
\label{sec:PnG}

Given the nature of antisymmetric correlations and the fact that it is effectively a squeezed bispectrum, it is natural to include the effect of primordial non-Gaussianity (PnG) in its expression.
Information about PnG is encoded within higher order correlations beyond the two-point function, pinpointing interactions of the inflaton with itself or with other fields and thus effectively acting as a particle detector for inflation~\cite{Bartolo2004, Chen2010, Wang2013}. Single-field models generate negligible non-Gaussianity, whereas multi-field inflationary scenarios provide in general larger values; for this reason, a measurement of $\fnl>1$ would rule out most single-field inflation models~\cite{Alvarez2014, dePutter2016,Byrnes2010}. The first statistics that is sensitive to deviations from Gaussianity is the bispectrum: its shape dependence is determined by the physical mechanisms at work during inflation, the most common templates being the local shape that peaks on squeezed triangles $\kl \ll k_1,k_2$, the equilateral shape $k_1 \sim k_2 \sim \kl$, and the folded shape $k_1 \sim k_2 \sim \kl/2$~\cite{Babich2004}.

CMB measurements from the Planck satellite constrain $\fnl^{\rm local} = -0.9 \pm 5.1$, $\fnl^{\rm equil} = -26 \pm 47$, $\fnl^{\rm ortho} = -38 \pm 24$ at 68\% C.L.~\cite{Planck2019}, and improvements are generally expected from large-scale structure measurements. For a review on methodologies and constraints on non-Gaussianity using power-spectrum and bispectrum measurements, see~\cite{Slosar2008,Ross2012,Agarwal2013} and~\cite{Karagiannis2018} respectively, and references therein. More recently,~\cite{Cabass2022b} established $\fnl^{\rm local} = -33 \pm 28$ at 68\% C.L. from the BOSS data, a constraint that is expected to improve significantly as future surveys such as SPHEREx~\cite{Dore2014}, which will access larger cosmological volumes. See also~\cite{Raccanelli2014, dePutter2014} for a discussion on the optimization of surveys to measure PnG.

It is worth noticing that even with just linear bias, non-Gaussianity can source an antisymmetric signal~\cite{Dai2015}. If $\langle\delta_{\tA}(\vec{k}_1)\delta_{\tB}(\vec{k}_2)\delta(\veckl)\rangle=\bltA(k_1)\bltB(k_2)B_{\rm grav}(\vec{k}_1,\vec{k}_2,\veckl)$, then
\begin{equation}
\label{eq:fLfNL}
    f_L^{{\rm A},\fnl} \simeq \fnl (\bltA-\bltB) \frac{ 3 \delta_{\rm cr} \Omega_{\rm m,0}H_0^2 \kl}{2 D_{\rm md}(z) T(\ks) \ks^4} \frac{ B_{\rm grav}(\vec{k}_1,\vec{k}_2,\veckl) }{P(\kl)} \, .
\end{equation}
The bulk of the information for this signal is in the long wavelength modes, then the amplitude decreases for smaller scales, faster than the biased clustering one.

As an indicative example of what could happen for other shapes and early Universe models, alternative scale-dependent bias behaviors are presented, that are taken to roughly model what would happen for other shapes of PnG.
A scale-dependent modifications of the bias $\propto \Delta b_1 k^n$, with $n=\{-2, -1, 0\}$ could mimic local, orthogonal, and equilateral shapes respectively~\cite{Raccanelli2015}.
In general, if $b_1^{X}(k)=b_1^{X}+\fnl\Delta b_1^{X} k^n$, then
\begin{equation}
    f_L^{{\rm A},\fnl} \simeq \frac{1}{2} \fnl \left(\bltA \Delta\bltB-\bltB \Delta\bltA\right) \frac{n \ks^n \kl}{2 \ks^2} \frac{ B_{\rm grav}(\vec{k}_1,\vec{k}_2,\veckl) }{P(\kl)} \, .
\end{equation}
The local PnG expression $n=-2$ is recovered for $\Delta b_1^X k^n = 3\delta_{\rm cr}(b_1^X-1)\Omega_{\rm m,0} H_0^2 / \left[ D_{\rm md}(z) T(k) k^2 \right]$.

An equilateral shape would leave no signature.

This estimate for $f_L^{\rm A,\fnl}$ is obtained by approximating the bispectrum as being simply proportional to the gravitational contribution, via the two linear bias parameters, $\langle\delta_{\tA}(\vec{k}_1)\delta_{\tB}(\vec{k}_2)\delta(\veckl)\rangle=\bltA(k_1)\bltB(k_2)B_{\rm grav}(\vec{k}_1,\vec{k}_2,\veckl)$. However, this does not capture all the ingredients: the signal needs to be modeled using the full second order kernels.

In the presence of local-type PnG, the Eulerian basis of operators in the bias expansion is modified by additional terms~\cite{Desjacques2016}, which at linear level give rise to the well-known scale-dependent bias~\cite{Matarrese2008, Dalal2007, Slosar2008, Verde2009, Desjacques2010, Xia2010}. Let $\phi$ be the Bardeen gravitational potential, related to the primordial curvature perturbation by $\phi=(3/5)\mathcal{R}$ and to the linear density field by
\begin{equation}
    \delta(\vec{k},\tau) = \mathcal{M}(k,\tau) \phi(\vec{k}) \, , \qquad\qquad \mathcal{M}(k,\tau) = \frac{2}{3} \frac{k^2T(k) D_{\rm md}(z)}{\Omega_{\rm m,0} H_0^2} \, ,
\end{equation}
where $D_{\rm md}(z)$ is the linear growth factor normalized to $a$ during matter domination.

The new operators are $\fnl\phi(\vec{q})$ at first order and $\fnl\delta(\vec{x})\phi(\vec{q})$ at second order. The Bardeen potential is evaluated at the Lagrangian position $\vec{q}$, since the coupling between non-Gaussianity of primordial fluctuations and matter fluctuations is imprinted in the initial conditions, and not induced by evolution. Therefore, these operators get contributions from the (linear-order) displacement field $\vec{s}$, given by
\begin{equation}
	\vec{x}=\vec{q}+\vec{s}(\vec{q},\tau) \, , \qquad\qquad \vec{s}(\vec{q},\tau) = -\frac{\nabla}{\nabla^2}\delta(\vec{q},\tau) \, .
\end{equation}
The contributions from PnG are
\begin{equation}
    \left.\delta_g(\vec{x})\right|_{\fnl} = \fnl b_{\phi} \phi(\vec{q}) +\fnl b_{\phi\delta}\phi(\vec{q})\delta(\vec{x}) \, .
\end{equation}
When adding RSD, the new terms arise from $\delta$ and $\partial_r(\delta v)$: any effect of $\fnl$ on velocities can be neglected, because it would be of the same order as higher derivative operators of the form $\partial^2\phi$~\cite{Cabass2022a}. Including PnG, thus, the new contributions to the redshift space kernels in Fourier space are
\begin{equation}
    Z_{1,\fnl}(\vec{k}) = \fnl b_{\phi} \mathcal{M}^{-1}(k) \, ,
\end{equation}
\begin{equation}
\begin{split}
    Z_{2,\fnl}(\vec{p},\vec{q}) =& \fnl b_{\phi} \frac{\vec{p}\cdot\vec{q}}{2} \left( \frac{1}{p^2}\mathcal{M}^{-1}(q) +\frac{1}{q^2}\mathcal{M}^{-1}(p) \right) +\fnl b_{\phi\delta} \frac{1}{2}\left(\mathcal{M}^{-1}(p)+\mathcal{M}^{-1}(q)\right) \\
    & +\fnl b_{\phi} \frac{k\mu f}{2} \left( \frac{\mu_q}{q}\mathcal{M}^{-1}(p) +\frac{\mu_p}{p}\mathcal{M}^{-1}(q) \right) \, .
\end{split}
\end{equation}

\begin{figure}[h!]
    \centering
    \includegraphics[width=.5\textwidth]{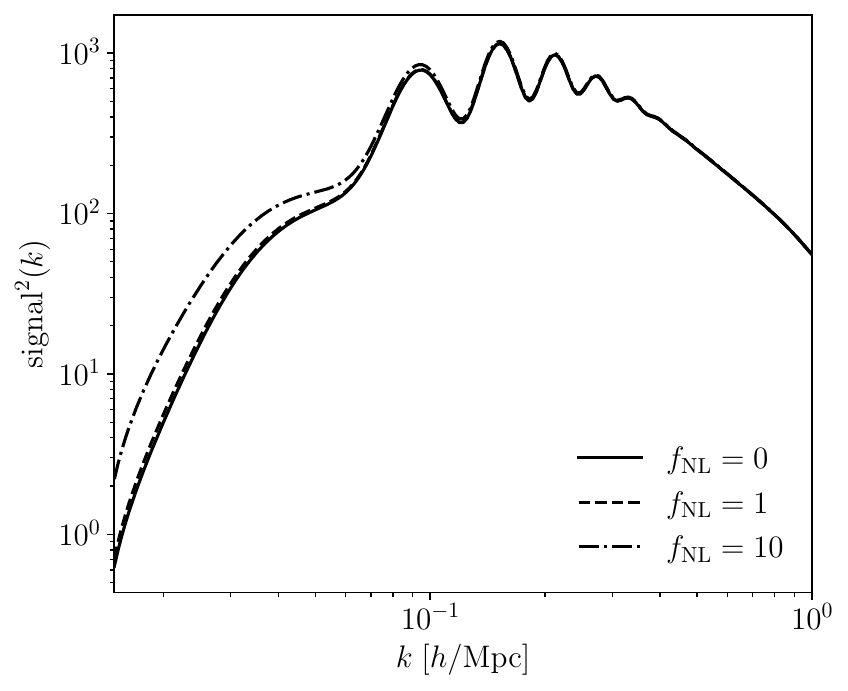}
    \caption{Integral over long modes of $\left(\PA\right)^2$, with biases $\bltA=3$, $\bltB=1.1$. The signal is modeled using full kernels and local type PnG, as in Equation~\eqref{eq:fullexprFK}. This assumes a survey with volume $V_s = 100 \ ({\rm Gpc}/h)^3$, centered at $z=1$, and galaxy number densities $n_{g,A}=10^{-3} \ (h/\Mpc)^3$ and $n_{g,B}=2n_{g,A}$.}
    \label{fig:signal2_fNL}
\end{figure}

Working at linear order in $\fnl$, the additional contribution to $\langle \delta_{\tA}(\vec{k}_1)\delta_{\tB}(\vec{k}_2)\delta(\veckl) \rangle$ coming from primordial non-Gaussianity is
\begin{equation}
\begin{split}
    2 \fnl &\Big\{ P(k_1)P(k_2) \Big( Z_{1,\fnl}^{\tA}(\vec{k}_1) Z_1^{\tB}(\vec{k}_2) Z_2(-\vec{k}_1,-\vec{k}_2) \\
    &\qquad +Z_1^{\tA}(\vec{k}_1) Z_{1,\fnl}^{\tB}(\vec{k}_2) Z_2(-\vec{k}_1,-\vec{k}_2) \\
    &\qquad +Z_1^{\tA}(\vec{k}_1) Z_1^{\tB}(\vec{k}_2) Z_{2,\fnl}(-\vec{k}_1,-\vec{k}_2) \Big) \\
    & +P(k_1) P(\kl) \Big( Z_{1,\fnl}^{\tA}(\vec{k}_1) Z_2^{\tB}(-\vec{k}_1,-\veckl) Z_1(\veckl) \\
    &\qquad +Z_1^{\tA}(\vec{k}_1) Z_2^{\tB}(-\vec{k}_1,-\veckl) Z_{1,\fnl}(\veckl) \\
    &\qquad +Z_1^{\tA}(\vec{k}_1) Z_{2,\fnl}^{\tB}(-\vec{k}_1,-\veckl) Z_1(\veckl) \Big) \\
    & +P(k_2) P(\kl) \Big( Z_2^{\tA}(-\vec{k}_1,-\veckl) Z_{1,\fnl}^{\tB}(\vec{k}_2) Z_1(\veckl) \\
    &\qquad +Z_2^{\tA}(-\vec{k}_2,-\veckl) Z_1^{\tB}(\vec{k}_2) Z_{1,\fnl}(\veckl) \\
    &\qquad +Z_{2,\fnl}^{\tA}(-\vec{k}_2,-\veckl) Z_1^{\tB}(\vec{k}_2) Z_1(\veckl) \Big) \Big\} \, .
\end{split}
\end{equation}
Antisymmetrizing and taking the squeezed limit, the contributions from non-Gaussianity effects are of order $(\kl/\ks)^{-1}$, $(\kl/\ks)^0$, $(\kl/\ks)^1$, and higher order. In the squeezed configuration, the first contribution is expected to be the dominant one
\begin{equation}
\begin{split}
    \frac{ B^{{\rm A},\fnl}(\vecks,\veckl) }{ P(\kl) (1+f\mul^2) } =& \frac{3 \fnl H_0^2\Omega_{\rm m,0}}{4 \ks \kl D_{\rm md}(z)} P(\ks) \\
    & \Bigg\{ \bphi f \mus^2 \left[ -2 \left( \musl +2 f \mus^2 \musl -f \mus \mul \right) +f \mus^2 \musl \dlogPkdlogk \right] \\
    & +\bphidelta f \mus \left[ -2 \left( \mus \musl -\mul \right) +\mus \musl \dlogPkdlogk \right] \\
    & +\blphi \musl \left[ 2(1+f\mul^2) -f \mus^2 \dlogPkdlogk \right] \\
    & -\blphidelta \musl \dlogPkdlogk \Bigg\} \, .
\end{split}
\end{equation}
This is the leading $\fnl$ contribution to the antisymmetric signal from nonlinear biased clustering. The signal, integrated over long modes and over angles, is shown in Figure~\ref{fig:signal2_fNL}.

Going beyond local PnG requires the introduction of an additional bias operator~\cite{Desjacques2016}
\begin{equation}
    \psi(\vec{q}) = \int\frac{d^3\vec{k}}{(2\pi)^3} k^{\alpha}\phi(\vec{k}) e^{i\vec{k}\cdot\vec{q}} \, ,
\end{equation}
again evaluated at the Lagrangian position $\vec{q}$. The resulting contributions to the redshift space kernels in Fourier space are
\begin{equation}
    Z_{1,\rm PnG}(\vec{k}) = a_0 b_{\psi} k^{\alpha} \mathcal{M}^{-1}(k) \, ,
\end{equation}
\begin{equation}
\begin{split}
    Z_{2,\rm PnG}(\vec{p},\vec{q}) =& a_0 b_{\psi} \frac{\vec{p}\cdot\vec{q}}{2} \left( \frac{1}{p^2}q^{\alpha}\mathcal{M}^{-1}(q) +\frac{1}{q^2}p^{\alpha}\mathcal{M}^{-1}(p) \right) +a_0 b_{\psi\delta} \frac{1}{2}\left(p^{\alpha}\mathcal{M}^{-1}(p)+q^{\alpha}\mathcal{M}^{-1}(q)\right) \\
    & +a_0 b_{\psi} \frac{k\mu f}{2} \left( \frac{\mu_q}{q}p^{\alpha}\mathcal{M}^{-1}(p) +\frac{\mu_p}{p}q^{\alpha}\mathcal{M}^{-1}(q) \right) \, .
\end{split}
\end{equation}
The local PnG case discussed above is recovered by setting $\alpha=0$, $a_0b_{\psi}=\fnl b_{\phi}$, and more details on the PnG bias parameters can be found in Appendix~\ref{app:biases}.
Detectability prospects for the local PnG case will be discussed in Section~\ref{sec:PnGconstraints}. A more detailed analysis of the antisymmetric signal and of the detectability prospects for the orthogonal and folded shapes of PnG is of great interest and left for a future work.

\section{Antisymmetric galaxy correlation in redshift-space, including primordial non-Gaussianity}

Here is the full expression, calculated for the first time, for the antisymmetric correlation in redshift-space, including second order bias and primordial non-Gaussianity.

\begin{equation}
\label{eq:fullexprFK}
\begin{split}
    & \frac{ B^{\rm A}(\vec{k}_1,\vec{k}_3) }{ P(k_3) (1+f\mu_3^2) } = \frac{1}{42} \Koverk P(\ks) \Bigg\{ 3\bll f\mu_1 \Bigg[ \dlogPkdlogk \mu_{13} \left(7f\mu_1^3-14\mu_{13}\mu_3+\mu_1 (4+10\mu_{13}^2-7f\mu_3^2)\right) \\
    & \qquad\qquad +2 \left(-14 f\mu_1^3\mu_{13}+7f\mu_1^2\mu_3+3(3+4\mu_{13}^2)\mu_3+\mu_1\mu_{13}(-13-8\mu_{13}^2+7f\mu_3^2)\right) \Bigg] \\
    & \qquad +21\bss f\mu_1 \left( \dlogPkdlogk \mu_1\mu_{13} +2(-\mu_1\mu_{13}+\mu_3) \right) \\
    & \qquad +14\bKK f\mu_1 \left(\dlogPkdlogk \mu_1\mu_{13} (3\mu_{13}^2-1) -2(-4 \mu_1\mu_{13}+6\mu_1\mu_{13}^3+\mu_3-3\mu_{13}^2\mu_3)\right) \\
    & \qquad -21 \bls \dlogPkdlogk \mu_{13} \\
    & \qquad +14 \blK \mu_{13} \left( 6(\mu_{13}^2-1) +\dlogPkdlogk (1-3\mu_{13}^2) \right) \Bigg\} \\
    & +\frac{3\fnl H_0^2\Omega_{\rm m,0}}{4 D_{\rm md}(z) \ks^2 T(\ks)} \frac{1}{42} P(\ks) \Bigg\{ \bphi \left[ \Koverk^{-1}\mathcal{F}_{\phi}^{(-1)}(\vecks,\veckl) +\mathcal{F}_{\phi}^{(0)}(\vecks,\veckl) +\Koverk\mathcal{F}_{\phi}^{(1)}(\vecks,\veckl) \right] \\
    & \qquad +\bphidelta \left[ \Koverk^{-1}\mathcal{F}_{\phi\delta}^{(-1)}(\vecks,\veckl) +\mathcal{F}_{\phi\delta}^{(0)}(\vecks,\veckl) +\Koverk\mathcal{F}_{\phi\delta}^{(1)}(\vecks,\veckl) \right] \\
    & \qquad +\blphi \left[ \Koverk^{-1}\mathcal{F}_{1,\phi}^{(-1)}(\vecks,\veckl) +\mathcal{F}_{1,\phi}^{(0)}(\vecks,\veckl) +\Koverk\mathcal{F}_{1,\phi}^{(1)}(\vecks,\veckl) \right] \\
    & \qquad +\blphidelta \left[ \Koverk^{-1}\mathcal{F}_{1,\phi\delta}^{(-1)}(\vecks,\veckl) +\mathcal{F}_{1,\phi\delta}^{(0)}(\vecks,\veckl) +\Koverk\mathcal{F}_{1,\phi\delta}^{(1)}(\vecks,\veckl) \right] \\
    & \qquad +\bphis \Koverk \mathcal{F}_{\phi,2}^{(1)}(\vecks,\veckl) \\
    & \qquad +\bphiK \Koverk \mathcal{F}_{\phi,K^2}^{(1)}(\vecks,\veckl) \Bigg\} \, .
\end{split}
\end{equation}
The $\mathcal{F}(\vecks,\veckl)$ functions are reported in Appendix~\ref{app:Fs}. They depend on the angles between long and short modes, and with the line of sight, and on derivatives of the power spectrum and of the transfer functions.

In the remainder of this work, the second order bias parameters will be obtained from the linear order ones by means of fitting formulae~\cite{Chan2012,Baldauf2012,Desjacques2016}, as reported in Appendix~\ref{app:biases}.

\renewcommand{\ks}{k}
\renewcommand{\kl}{K}
\renewcommand{\vecks}{\vec{k}}
\renewcommand{\veckl}{\vec{K}}
\newcommand{\ksmin}{k_{\rm min}}
\newcommand{\ksmax}{k_{\rm max}}
\newcommand{\klmin}{K_{\rm min}}
\newcommand{\klmax}{K_{\rm max}}

\section{Detectability}
\label{sec:detectability}

In~\cite{Jeong2012} and~\cite{Dai2015}, an estimator was built in order to extract information on the amplitude of the modulating long-wavelength field. The procedure aims to find the minimum detectable amplitude of the power spectrum of the underlying field, in the spirit of seeking for fossil imprints from primordial exotic physics. In principle, this could be generalized to the case of nonlinear biased clustering as well, and the calculation is reported in Appendix~\ref{app:estimator}. However, for practical purposes, one can take a different approach and build a signal-to-noise ratio estimator, in the same fashion as in~\cite{McDonald2009}.

Let $\ks$, $\kl$ be the short and long mode, respectively. Defining the antisymmetric signal as
\begin{equation}
    \PA(\vecks,\veckl) \equiv \frac{1}{2} \left[ \delta_{\tA}(\vecks)\delta_{\tB}(\veckl-\vecks) -\delta_{\tA}(\veckl-\vecks)\delta_{\tB}(\vecks) \right] \, ,
\end{equation}
at fixed long mode $\veckl$, the covariance in the null hypothesis is
\begin{equation}
    {\rm Cov}(\vecks,\vecks\pr)_{\veckl} = \frac{1}{2} \left[ P_{\tA\tA}(\ks)P_{\tB\tB}(\ks) -P_{\tA\tB}(\ks)P_{\tA\tB}(\ks) \right] \left[ \delta^D_{\vecks+\vecks\pr} -\delta^D_{\vecks-\vecks\pr} \right] \, .
\end{equation}
This matrix is not invertible; however, following the same steps as~\cite{Zhou2020}, one can notice that $\PA(-\vec{k})=-\PA(\vec{k})$, and therefore consider only one emisphere in $\vec{k}$ space and combine the contribution from both $\vec{k}$ and $-\vec{k}$ mode
\begin{equation}
    \left[ \langle \PA(\vec{k})^2 \rangle-\langle \PA(\vec{k}) \rangle^2 \right] - \left[ \langle \PA(\vec{k}) \PA(-\vec{k}) \rangle-\langle \PA(\vec{k}) \rangle \langle \PA(-\vec{k}) \rangle \right] \, .
\end{equation}
The covariance on the emisphere is
\begin{equation}
\label{eq:covemi}
    {\rm Cov}^{\rm emi}(\vecks,\veckl) = \frac{1}{2} \Big[ P_{\tA\tA}(\ks)P_{\tB\tB}(|\veckl-\vecks|) +P_{\tA\tA}(|\veckl-\vecks|)P_{\tB\tB}(\ks) \Big] -P_{\tA\tB}(\ks)P_{\tA\tB}(|\veckl-\vecks|) \, .
\end{equation}

The signal depends on the short mode $\vecks$, the long mode $\veckl$, and the angle between them.
First, one has to integrate over the long modes: in this work, the long modes and short modes are chosen in such a way that they are separated by at least an order of magnitude, that is, the minimum squeezing factor is set to $10$, so that there is a hierarchy $\kl < \klmax \ll \ksmin < \ks$.
As anticipated in Section~\ref{sec:RSD}, the SNR unavoidably depends on the arbitrary choice of squeezing factor\footnote{Compare e.g.,~to the situation described in~\cite{dePutter2018}, where the bulk of the information is contained in the most squeezed configurations, and therefore the analysis can be pushed to $\klmax=\ksmin$, including triangles that are technically not squeezed at all.}. Of course, the smaller the squeezing factor, the more triangular configurations will enter in the integral, and the larger the SNR; but, on the other hand, the formal derivation has been made in the $\kl \ll \ks$ limit, therefore one should be careful in including configurations that are not ``squeezed enough''. There has been recently a study in the context of intensity mapping reconstruction~\cite{Wang2023} that investigates how to deal with this issue by keeping higher order terms. While that approach will be useful when trying to extract the full signal in the antisymmetric case, for the current purposes here a simpler procedure will be adopted, adding by hand a squeezing factor.

Each pair of short modes $\vec{k}_1$, $\vec{k}_2$ is sensitive to the modulation induced by a long mode $\veckl=\vec{k}_1+\vec{k}_2$, therefore the minimum accessible long mode $\klmin$ is taken to be of the order the fundamental wavenumber of the survey $\kf=2\pi/V_s^{1/3}$.
In order to restrict to squeezed configurations only, for a given short mode $\ks$ the corresponding maximum long mode is $\klmax = \ks/10$.
The short modes will span from $\ksmin \gtrsim 10 \kf$ to some $\ksmax$.

The angles between the line-of-sight direction and the short wavelength wavevector $\vec{k}_1$ are described by $\mu_1=\cos\theta_1$ and $\phi_1$, while the ones describing the long wavelength wavevector $\veckl$ are $\mu=\cos\theta$ and $\phi$.


The signal is $\PA(\vecks,\veckl) = \left( B^{\rm A}(\vecks,\veckl) / P(\veckl) \right) \delta(\veckl)$. When including RSD, the long mode power spectrum is $P(\veckl)=\left(1+f\mul^2\right)P(\kl)$. The SNR$(\ks)$ is obtained integrating the signal squared $S^2=(\PA)^2$ over the noise $N^2={\rm Cov}^{\rm emi}$, on the corresponding range of long modes for each $\ks$, that is
\begin{equation}
    \left(\frac{S}{N}\right)_{\vecks}^2 = \int_{\kf}^{\ks/10} \frac{ \kl^2 d\kl }{(2\pi)^3} \int_{-1}^{+1} d\mu \int_0^{2\pi} d\phi \frac{1}{ {\rm Cov}^{\rm emi}(\vecks,\veckl) } \left( \frac{B^{\rm A}(\vecks,\veckl)}{P(\veckl)} \right)^2 P(\veckl) \, .
\label{eq:snrk}
\end{equation}
Then, integrating over short modes as well,
\begin{equation}
    \left(\frac{S}{N}\right)^2
    = \frac{V_s}{(2\pi)^3} \int_{\kf}^{\ksmax} \ks^2d\ks \int_0^{2\pi} d\phi_1 \int_0^{1} d\mu_1 \left\{ \int_{\kf}^{\ks/10} \frac{ \kl^2 d\kl }{(2\pi)^3} \int_{-1}^{+1} d\mu \int_0^{2\pi} d\phi \frac{1}{ {\rm Cov}^{\rm emi}(\vecks,\veckl) } \left( \frac{B^{\rm A}(\vecks,\veckl)}{P(\veckl)} \right)^2 P(\veckl) \right\} \, ,
\label{eq:snr}
\end{equation}
where now $\ksmin=\kf$ is set by the volume of the survey and $\ksmax$ is the minimum scale for the analysis. Notice that the integration over angles covers only the emisphere in $\vec{k}$-space.

\begin{figure}[h!]
    \centering
    \includegraphics[width=\textwidth]{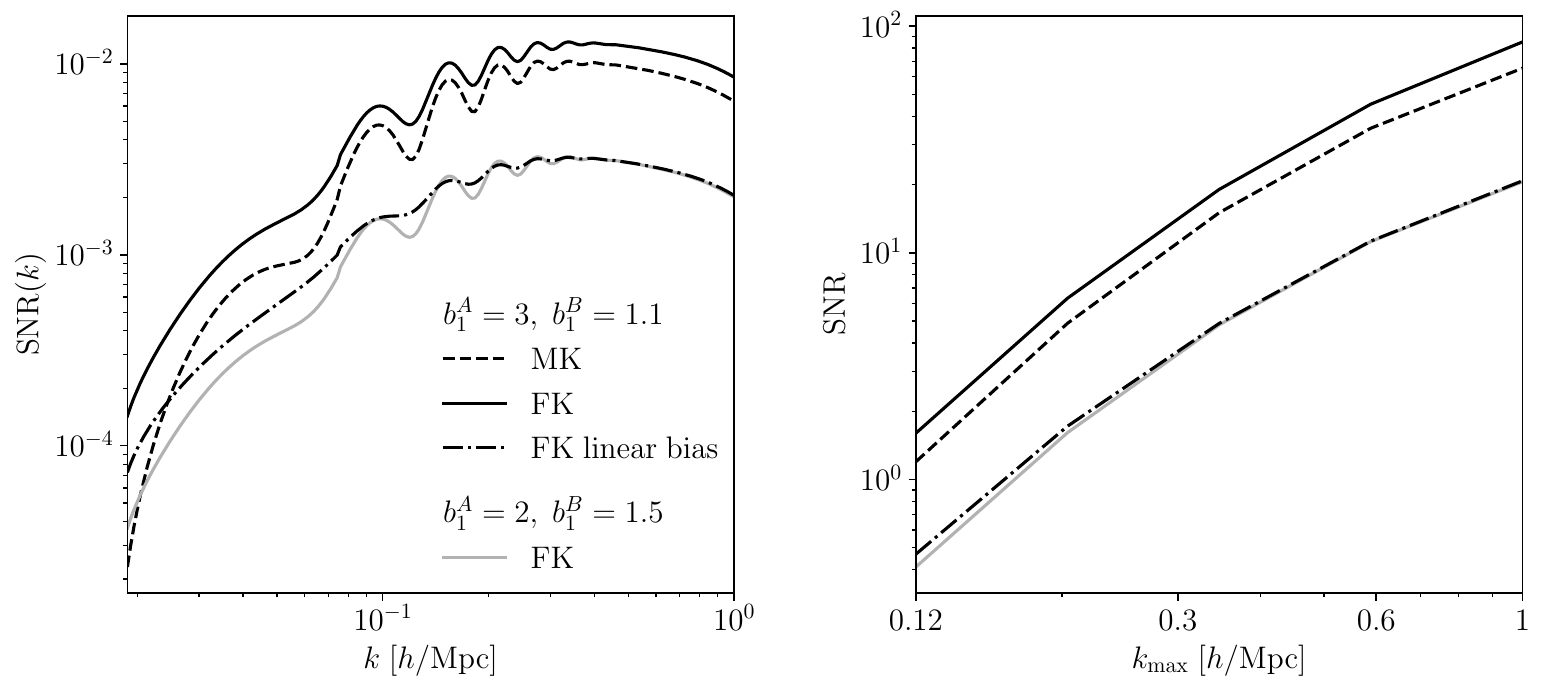}
    \caption{SNR for nonlinear biased clustering, sourced by the squeezed bispectrum between two tracers and an underlying dark matter long mode, antisymmetrized with respect to the two tracers. \textit{Left panel}: the SNR is plotted as a function of the short mode $k$ after integrating over long modes, Equation~\eqref{eq:snrk}. \textit{Right panel}: the SNR has been integrated over short modes as well, Equation~\eqref{eq:snr}, and it is plotted as a function of the maximum wavenumber of the analysis, $k_{\rm max}$.
    The various lines represent calculations with minimal kernels as in~\cite{Dai2015} (MK, dashed) and full kernels as computed in Section~\ref{sec:RSD} (FK, solid). The case for full kernels with linear bias only is also shown (dot-dashed).
    The plots show results for two example cases ($\bltA=3$, $\bltB=1.1$) in black, and ($\bltA=2$, $\bltB=1.5$) in gray.
    The survey volume is taken to be $50 \ ({\rm Gpc}/h)^3$, centered at $z=1$, with tracers density $n_{g,\tA}=10^{-3} \ (h/\Mpc)^3$, $n_{g,\tB}=2n_{g,\tA}$.}
    \label{fig:SNR_PA}
\end{figure}

Figure~\ref{fig:SNR_PA} shows the SNR for nonlinear biased clustering. It is shown both as a function of $\ks$ after integrating over the long modes $\kl$, and as a function of the maximum $\ksmax$ of the analysis after integrating over $\ks$ as well. Two cases are reported: the minimal kernel (MK in the plots) framework as in~\cite{Dai2015} and the full kernel (FK) one as described in this work. The minimal kernel assumption slightly underestimates the SNR with respect to the more accurate modeling.

A clever choice of the tracers can significantly boost the signal. For example, in the case of a $50 \ ({\rm Gpc}/h)^3$ survey, centered at $z=1$, for a choice of linear biases ($\bltA=2$, $\bltB=1.5$) an SNR of order $\mathcal{O}(10)$ is achieved at $k_{\rm max} \sim 0.6 \ h/\Mpc$. However, when targeting e.g.,~two populations with ($\bltA=3$, $\bltB=1.1$) the SNR is $\mathcal{O}(10)$ at $k_{\rm max} \sim 0.3 \ h/\Mpc$, which is within the reach of current analyses~\cite{DAmico2019, Baldauf2016, Chudaykin2020a}.
This can be used as a guidance for setting the required source targeting, the exact situation being different for specific surveys.

The purely linear bias case is also shown: in the presence of RSD, there is a nonvanishing signal, which is harder to detect unless reaching higher $k_{\rm max}$. The signal comes from the fact that in multi-tracer analyses there will be a mixed bias-RSD term that carries information on the velocity-density spectrum, in a similar fashion as unequal time correlations~\cite{Raccanelli2023a}.

\subsection{Constraints on the bias}
\label{sec:biasconstraints}

It comes natural to investigate if the antisymmetric correlation, as it is depending on combinations of the bias parameters that are different from the standard case, can constrain them (or a combination of) in a meaningful way.

In the minimal kernels case, the signal is proportional to the antisymmetrized combination of bias parameters $\PA \propto (b_{\tB}c_{\tA}-b_{\tA}c_{\tB}) \equiv \biascombination$. Using this simplified case for an estimate of the precision reachable in measuring $\biascombination$, then $\sigma_{\biascombination}/\biascombination \sim 1/{\rm SNR}$ (68 \% C.L.). When including RSD, gravitational evolution, and the other second-order bias parameters, the scaling becomes much more complex and the constraining power cannot be summarized in such a straightforward way.

As mentioned in the previous Section, the linear bias parameters $\{ \bltA, \bltB \}$ will be connected to the higher order ones by means of fitting formulae~\cite{Lazeyras2015, Desjacques2016}, and reported in Appendix~\ref{app:biases}.

If one were to consider the entire set of parameters, the main limiting factor for the constraining power on the bias parameters would be degeneracies (as expected, and as it happens for standard observables including loops), especially when accounting for all terms included in this work.
As can be immediately seen from Equation~\eqref{eq:fullexprFK}, the biases $\{ \bltA, \bltB, \bstA, \bstB, \bKtA, \bKtB \}$ combine in the final signal in a way that makes it difficult to disentangle the single contributions. There is virtually no constraining power on the full set of bias parameters, and even pushing the analysis to a more aggressive $k_{\rm max} \sim 0.6 \ h/\Mpc$, the constraints are not competitive with other available observables.

\begin{figure}[h!]
    \centering
    \includegraphics[width=.45\textwidth]{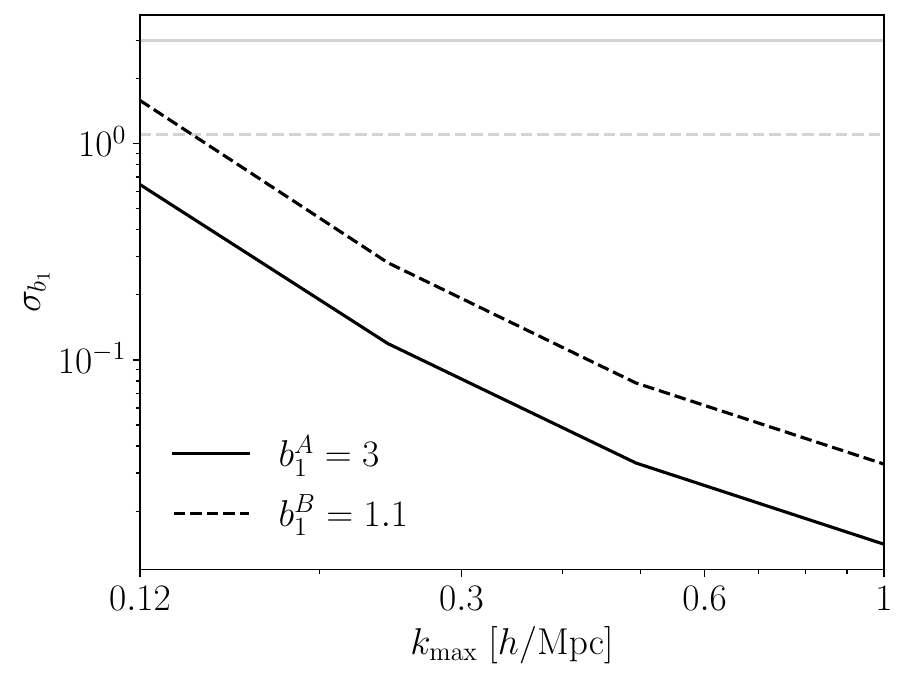}
    \includegraphics[width=.45\textwidth]{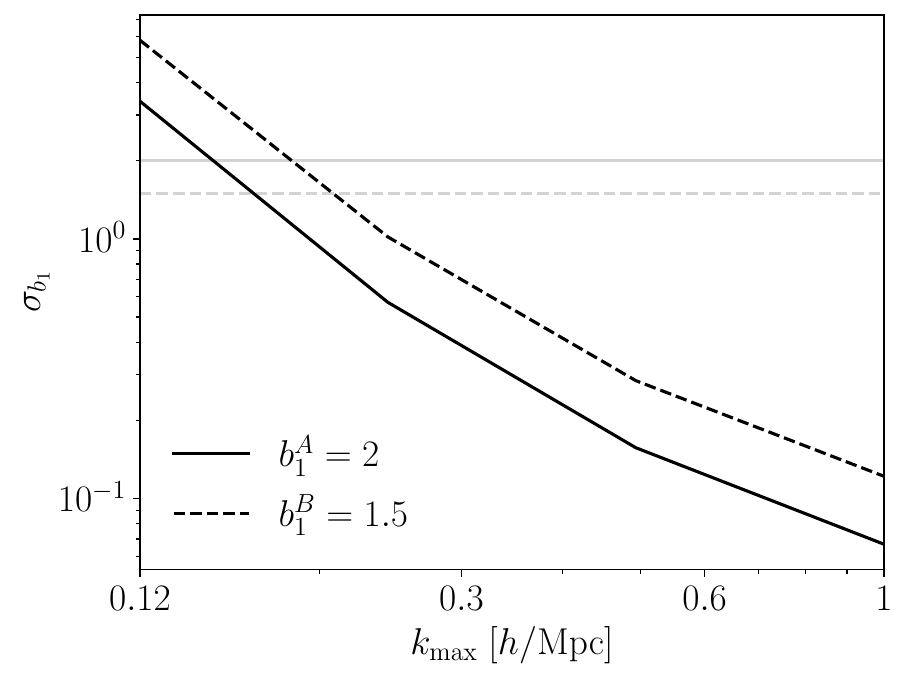}
    \caption{Forecasts on $\{ \bltA$, $\bltB \}$, with second order biases being linked to the linear ones by the fitting formulae. The plot shows the dependence on $k_{\rm max}$ of the $1\sigma$ constraints on $b_1^X$. The survey volume is $100 \ ({\rm Gpc}/h)^3$ centered at $z=1$, with tracers $n_{g,\tA}=10^{-3} \ (h/\Mpc)^3$, $n_{g,\tB}=2n_{g,\tA}$. The fiducial biases are set to ($\bltA=3$, $\bltB=1.1$) on the left panel, and to ($\bltA=2$, $\bltB=1.5$) on the right panel.
    }
    \label{fig:1sigma_b1only}
\end{figure}

Figure~\ref{fig:1sigma_b1only} shows the dependence of the constraining power on $\ksmax$ (where of course one would need to model non-linearities), for two choices of fiducial biases ($\bltA=3$, $\bltB=1.1$) and ($\bltA=2$, $\bltB=1.5$). Constraints of $\mathcal{O}(10-30\%)$ on the linear bias parameters are achieved when reaching $\ksmax \sim 0.3\ h/\Mpc$. This is not competitive with expected constraints from future galaxy surveys, but the analysis performed here is a very rough estimate: a more detailed procedure, including several redshift bins and their correlation and pushing the analysis to higher redshifts and smaller scales, may significantly improve the constraints.

\subsection{Primordial non-Gaussianity}
\label{sec:PnGconstraints}

This Section investigates the constraining power of the antisymmetric galaxy cross-correlation on local PnG. 
The signal is modeled using the full kernels, as in Equation~\eqref{eq:fullexprFK}, and assuming a fiducial $\fnl=1$ and $\{\bltA=3$, $\bltB=1.1 \}$.
Only the $\propto k^{-2}$ scale-dependent bias associated with local-type PnG is investigated here. As a first rough estimate of the constraining power, the results on $\sigma_{\fnl}$ are obtained via a Fisher forecast on $\fnl$ alone, with
\begin{equation}
    \sigma_{\fnl}^{-2} \simeq \frac{V_s}{(2\pi)^3} \int_{\kf}^{\ksmax} \ks^2d\ks \int_0^{2\pi} d\phi_1 \int_0^{1} d\mu_1 \left\{ \int_{\kf}^{\ks/10} \frac{ \kl^2 d\kl }{(2\pi)^3} \int_{-1}^{+1} d\mu \int_0^{2\pi} d\phi \frac{1}{{\rm Cov}^{\rm emi}(\vecks,\veckl)} \left( \frac{\partial P^{\rm A}(\vecks,\veckl)}{\partial\fnl} \right)^2 \right\} \, .
\end{equation}

A more complete analysis, including degeneracies -- especially with the linear bias parameters $\bltA$ and $\bltB$ -- and extending to orthogonal and folded shapes, is left for a future work.

\begin{figure}[h!]
    \centering
    \includegraphics[width=.45\textwidth]{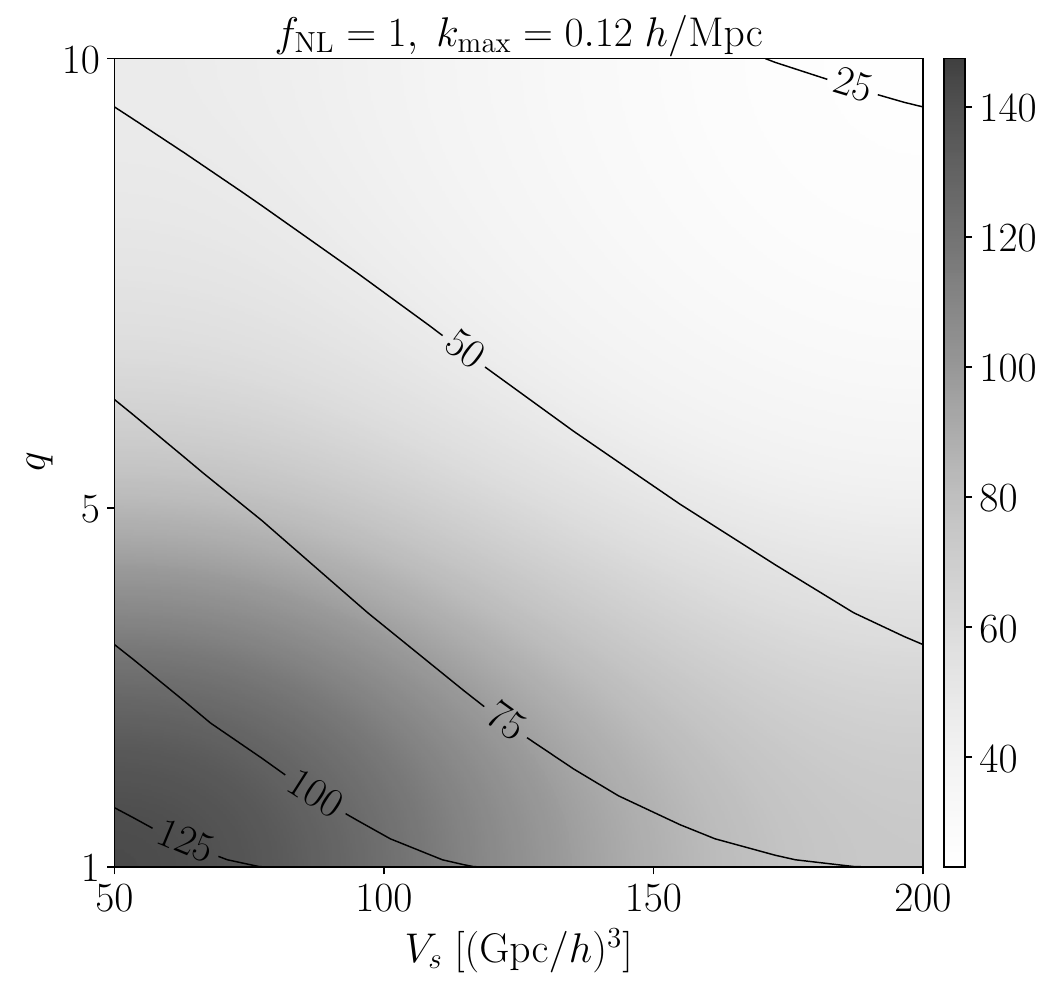}
    \includegraphics[width=.45\textwidth]{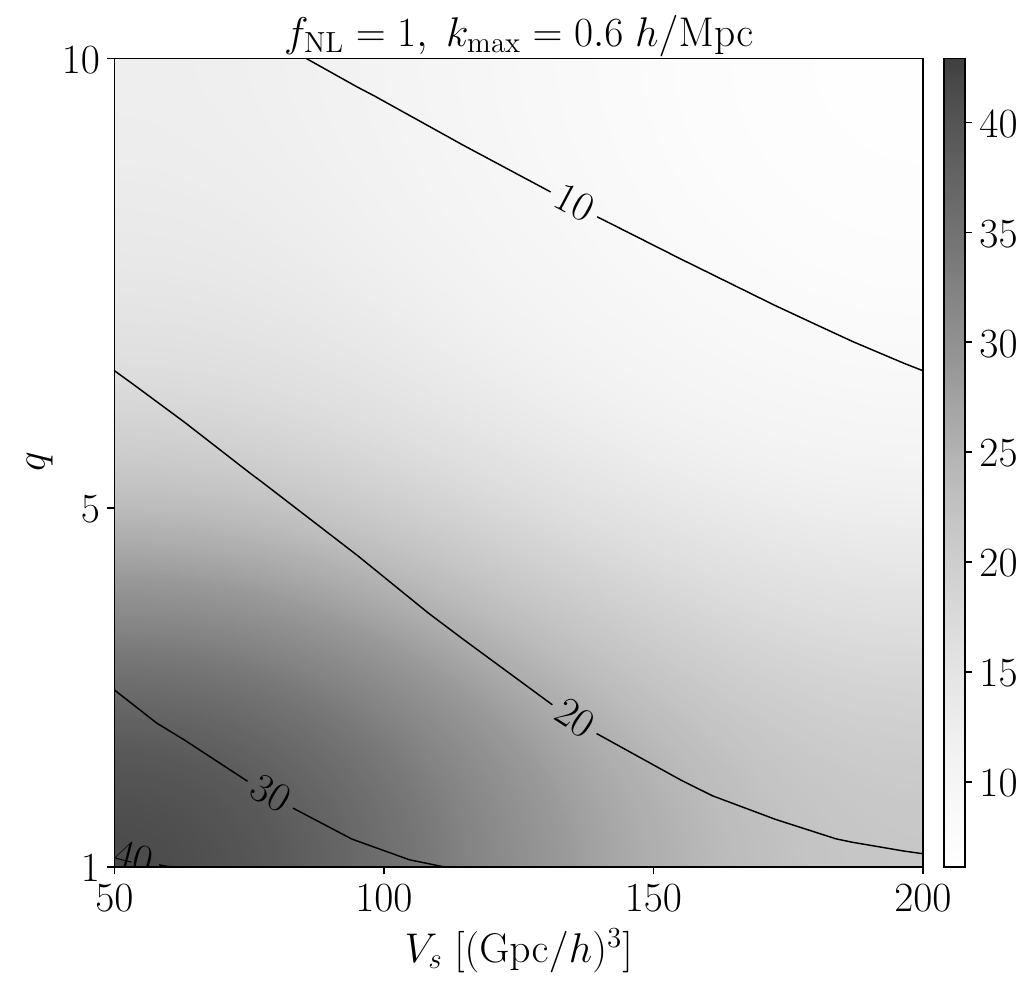}
    \caption{Value of $\sigma_{\fnl}$ when varying some survey specifications.
    }
    \label{fig:ContourPlot-sigmafNL}
\end{figure}

Figure~\ref{fig:ContourPlot-sigmafNL} investigates what would be the requirements for a survey to set stringent bounds on local $\fnl$ using this observable. Two directions are explored: a larger survey volume and a larger number density. Both are beneficial, but the improvement given by a larger survey volume tends to saturate: this is because, given the need to keep a hierarchy between long and short modes, the integration over short modes can never be pushed down to $k_{\rm min}=2\pi/V_s^{1/3}$.

The results fall in the same order of magnitude as recent constraints on $\fnl$ using the scale-dependent bias, obtained from the power spectrum of quasar samples in the eBOSS survey~\cite{Castorina2019,Mueller2021,Cabass2022b}, giving $1\sigma_{\fnl} \sim \mathcal{O}(20)$, as well as the forecasted constraining power for DESI and Euclid using the power spectrum only~\cite{Achucarro2022}, which will give similar results to the Planck ones $1\sigma_{\fnl} \sim \mathcal{O}(10)$. However, they are not competitive with a combined analysis of power spectrum and bispectrum, and with future radio~\cite{Ferramacho2014,Camera2015} and galaxy surveys such as SPHEREx~\cite{Dore2014} and the proposed SIRMOS~\cite{SIRMOS}, that forecast $\sigma_{\fnl}<1$.

\section{Tests of alternative cosmological models}
\label{sec:exoticmodels}

\renewcommand{\ks}{k_1}
\renewcommand{\kl}{k_3}
\renewcommand{\vecks}{\vec{k}_1}
\renewcommand{\veckl}{\vec{k}_3}

The antisymmetric galaxy cross-correlation is not just a new observable, but it presents the advantage of being sensitive to some exotic fundamental physics models that would otherwise not be tested when using more standard statistics; roughly, it can be thought of as something similar to tests performed with cosmic fossils~\cite{Dai2013a, Dimastrogiovanni2014}, but being sensitive to odd moments. The antisymmetric cross-correlation can be sensitive to models that include a preferred direction or where underlying long modes affect different populations in specific ways.

This Section investigates a few examples of beyond $\Lambda$CDM models that can imprint signatures that can be tested by the antisymmetric cross-correlation of two tracers.
Where not otherwise specified, the survey specifications are as in the previous Section.

\subsection{Vector modes}
\label{sec:vector}

In the presence of primordial vector fields from inflation~\cite{Bartolo2009, Dimastrogiovanni2010, Bartolo2015} -- for instance in axion inflation~\cite{Freese1990,Pajer2013} where the inflaton is coupled to a $U(1)$ gauge field, or due to primordial magnetic fields~\cite{Widrow2002,Subramanian2015} -- a vector polarization can arise in the presence of some preferred direction, to which the two tracers would respond in different ways, and therefore give a nonvanishing antisymmetric signal.

The generic parametrization of the antisymmetric component as in~\cite{Dai2015} is
\begin{equation}
\label{eq:genericparam}
    \PA(\vecks,\veckl) = \sum_p f_p^{\rm A}(\vecks,\veckl) h^*_p(\veckl) \hat{\epsilon}_p \cdot \left(2\vecks\musl+\veckl\right) \, ,
\end{equation}
with $p=L,x,y$. Taking the longitudinal polarization $\epsilon_L$ along the long mode direction, $\vhat{k}_3 \equiv \vhat{K}$, a complete orthonormal basis is given by
\begin{equation}
\begin{split}
    & \vhat{K} = \left( \cos\theta_3\cos\phi_3, \cos\theta_3\sin\phi_3, \sin\theta_3 \right) \, , \\
    & \vhat{X} = \left( \sin\phi_3, -\cos\phi_3, 0 \right) \, , \\
    & \vhat{Y} = \vhat{K}\times\vhat{X} = \left( \sin\theta_3\cos\phi_3, \sin\theta_3\sin\phi_3, -\cos\theta_3 \right) \, .
\end{split}
\end{equation}

\begin{figure}[htp!]
    \centering
    \includegraphics[width=.45\textwidth]{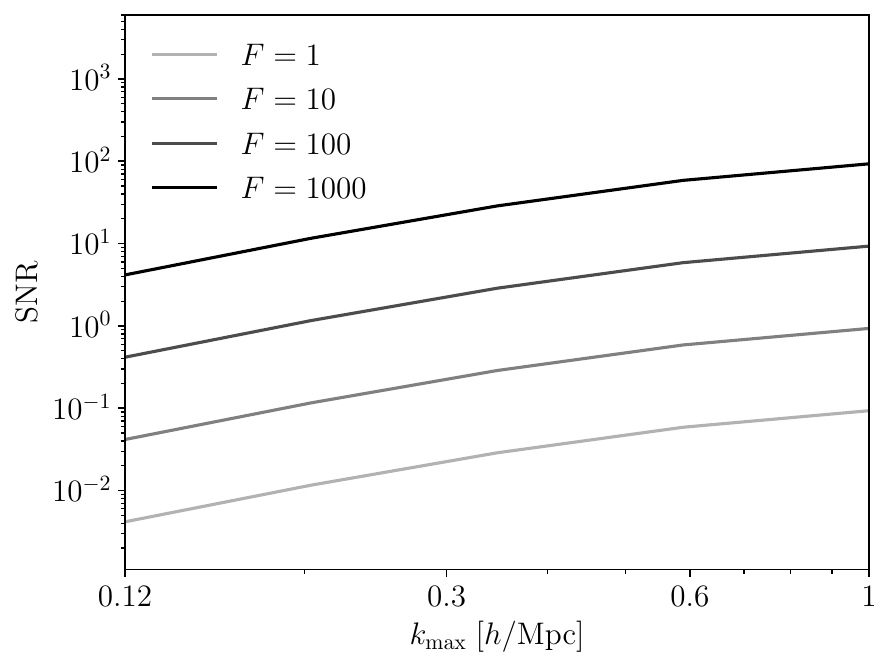}
    \caption{SNR for the two vector polarizations $p=x,y$, plotted as a function of $k_{\rm max}$, after integrating over both long and short modes. The long mode $h_p^*(\veckl)$ is a vector field of primordial origin, which power spectrum is taken to be a scale-invariant $F \left( 2\pi^2 A_s / \kl^3 \right)$; the SNR is shown for different amplitudes $F$.
    }
    \label{fig:SNR_vectormodes}
\end{figure}

Figure~\ref{fig:SNR_vectormodes} shows the SNR for a signal as in Equation~\eqref{eq:genericparam}, taking an amplitude~\cite{Jeong2012} $f^{\rm A}_{x,y}=-\frac{3}{2} Z_1^{\rm dm}(\vec{k}_1,z) P(k_1) / k_1$, where the dark matter kernel is $Z_1^{\rm dm}(\vec{k}_1,z)=\left(1+f\mu_1^2\right)$, and a scale-invariant power spectrum $P(\kl) = F \left( 2\pi^2 A_s / \kl^3 \right)$, with some amplitude $F$, which can be constrained.

The covariance is obtained assuming that there is no cross-correlation in the absence of the long mode, i.e.,~$2P_{\tA}^{\rm tot}(k_1)P_{\tB}^{\rm tot}(k_2)$, as in~\cite{Dai2015}.

\subsection{Two-component dark matter}
\label{sec:2dm}

There has recently been interest in models where the dark sector is not made by one single particle but it is extended and there are different particles in the dark sector, with particular focus on direct detection searches~\cite{Aguirre2004,Profumo2009,Adulpravitchai2011,Cyr-Racine2015,Herrero-Garcia2017,Herrero-Garcia2018}.
However, clustering properties of different biased tracers may provide an alternative detection method: it was proposed in~\cite{Dai2015} that a two-component dark matter model may leave an antisymmetric imprint if the two tracers cluster in different ways.

As an example, one can imagine that the dark matter components are affected by a relative modulation that imprints a preferred direction: for simplicity, let's associate it to the second tracer, $\delta_{\tB}(\vec{k})=\bltB\delta(\vec{k})\left(1+A\vhat{k}\cdot\vhat{p}\right)$, with $\vhat{p}$ the preferred direction and $A$ controlling the strength of the modulation. Then
\begin{equation}
    \langle \delta_{\tA}(\vec{k}_1) \delta_{\tB}(\vec{k}_2) \delta(\veckl) \rangle = \bltA\bltB\left(1+A\vhat{k}_2\cdot\vhat{p}\right) \langle\delta(\vec{k}_1)\delta(\vec{k}_2)\delta(\veckl)\rangle \, ,
\end{equation}
so that, after antisymmetrization, one is left with a signal $\propto \bltA\bltB B_{\rm grav}(k_1,k_2,\kl) A (\vhat{k}_2-\vhat{k}_1)\cdot\vhat{p}$ with $\vhat{\epsilon}=\vhat{p}$.
Adding RSD, in the case of generic kernels, the antisymmetrized bispectrum between the two tracers is
\begin{equation}
\label{eq:BA2dm}
    B^{\rm A}(\vec{k}_1,\vec{k}_2,\veckl) = \frac{1}{2} \Big[ \left( 1+A\vhat{k}_2\cdot\vhat{p} \right) \langle\delta_{s,\tA}(\vec{k}_1)\delta_{s,\tB}(\vec{k}_2)\delta_s(\veckl)\rangle - \left( 1+A\vhat{k}_1\cdot\vhat{p} \right) \langle\delta_{s,\tB}(\vec{k}_1)\delta_{s,\tA}(\vec{k}_2)\delta_s(\veckl)\rangle \Big] \, .
\end{equation}
This sources a signal
\begin{equation}
    \PA(\vec{k}_1,\veckl) = \left. \frac{ B^{\rm A}(\vec{k}_1,\vec{k}_2,\veckl) }{ P(\veckl) } \right|_{\vec{k}_1+\vec{k}_2+\veckl=0} \delta^*(\veckl) \, .
\end{equation}

Notice that, since RSD leave a nonvanishing signature even with linear bias only, this signal will get contributions that come from Equation~\eqref{eq:BAlinearbias} and are not due to the new ingredient $\propto A$.

The (emisphere) covariance will also get modified with respect to Equation~\eqref{eq:covemi} as
\begin{equation}
\begin{split}
    {\rm Cov}^{\rm emi}(\vec{k}_1,\vec{k}_2,\veckl) =& \frac{1}{2} \left[ (1+A\vhat{k}_2\cdot\vhat{p})^2 P_{\tA\tA}(k_1) P_{\tB\tB}(k_2) + (1+A\vhat{k}_1\cdot\vhat{p})^2 P_{\tA\tA}(k_2) P_{\tB\tB}(k_1) \right] \\
    & -(1+A\vhat{k}_1\cdot\vhat{p})(1+A\vhat{k}_2\cdot\vhat{p}) P_{\tA\tB}(k_1) P_{\tA\tB}(k_2) \, .
\end{split}
\end{equation}

\begin{figure}[h!]
    \centering
    \includegraphics[width=.5\textwidth]{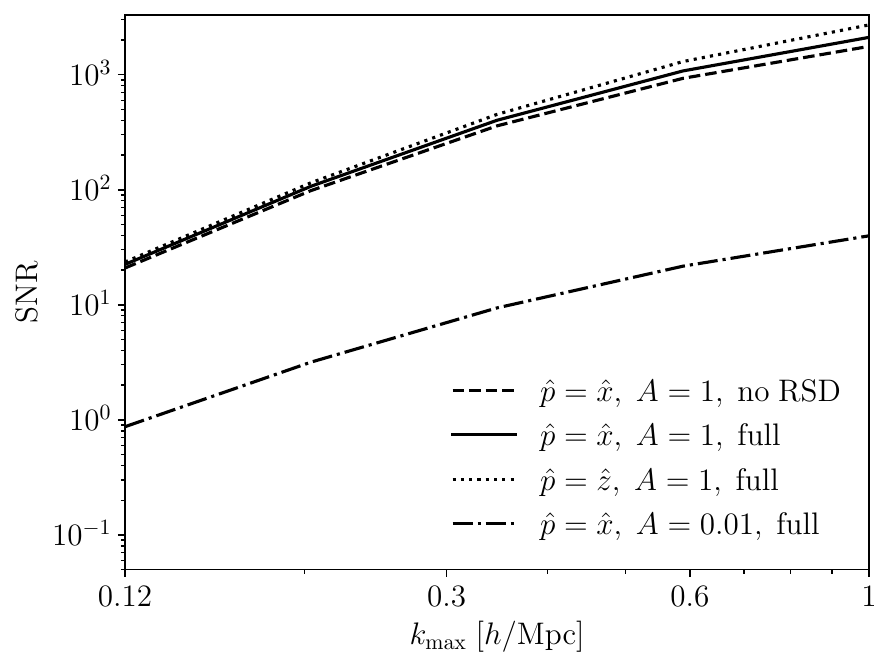}
    \caption{SNR for the amplitude of the two-component dark matter model, with amplitude $A$ and preferred direction either aligned with the line of sight $\vhat{p}=(0,0,1)$, or orthogonal to it, as an example $\vhat{p}=(1,0,0)$. A linear bias relation is assumed. The solid and dot-dashed lines include both RSD and gravitational evolution, while the dashed line shows the signal without RSD.}
    \label{fig:SNR2dm}
\end{figure}

Figure~\ref{fig:SNR2dm} shows the SNR for the signal above. The procedure of exchanging the two tracers is actually sensitive to the projection of the relative displacement between tracers onto the Fourier plane embedding the three wavevectors. Those wavevectors that happen to lie orthogonally to the preferred direction will give no contribution to the signal.

\subsection{Anisotropies from inflation}
\label{sec:anisotropies}

Statistical isotropy and/or homogeneity may be slightly violated, due to some physical process of primordial origin. In the presence of a dipolar or quadrupolar (or higher) modulation of the primordial curvature power spectrum, the matter power spectrum inherits a new factor so that antisymmetric correlations could test such a model. In this case, the power spectrum can be written as~\cite{Ackerman2007,Shiraishi2016}
\begin{equation}
    \langle\zeta(\vec{k}_1)\zeta(\vec{k}_2)\rangle = (2\pi)^3 \delta^{(3)}(\vec{k}_1+\vec{k}_2) P_{\zeta}(k_1) \left[ 1+\Qfenicio(\vec{k}_1) \right] \, ,
\end{equation}
where, for a generic distortion multipole $L$,
\begin{equation}
    \Qfenicio(\vec{k}_1) = \sum_M g_{LM} f(k_1) Y_{LM}(\vhat{k}_1) \, .
\end{equation}
with $g_{LM}$ being the amplitudes of the modulation and $f(k)$ carrying the scale dependence.
Notice that this case is different from the previous examples: here, the two tracers respond in the same way to the primordial modulation that is imprinted in the initial conditions. In this sense, the antisymmetric signal is still the one that was calculated for the nonlinear biased clustering, but the SNR is modified by the additional anisotropic factor that comes from the primordial curvature power spectrum. In fact, the extra factor can be interpreted as an effective scale-dependent bias, $\tilde{b}_1(k) = b_1\left(1+\Qfenicio(\vec{k})\right)$; however, the scale-dependence is the same for every tracer, and in the case of purely linear bias and no RSD, the antisymmetric signal would vanish, the same way it would for nonlinear clustering.

 \begin{figure}[h!]
     \centering
     \includegraphics[width=\textwidth]{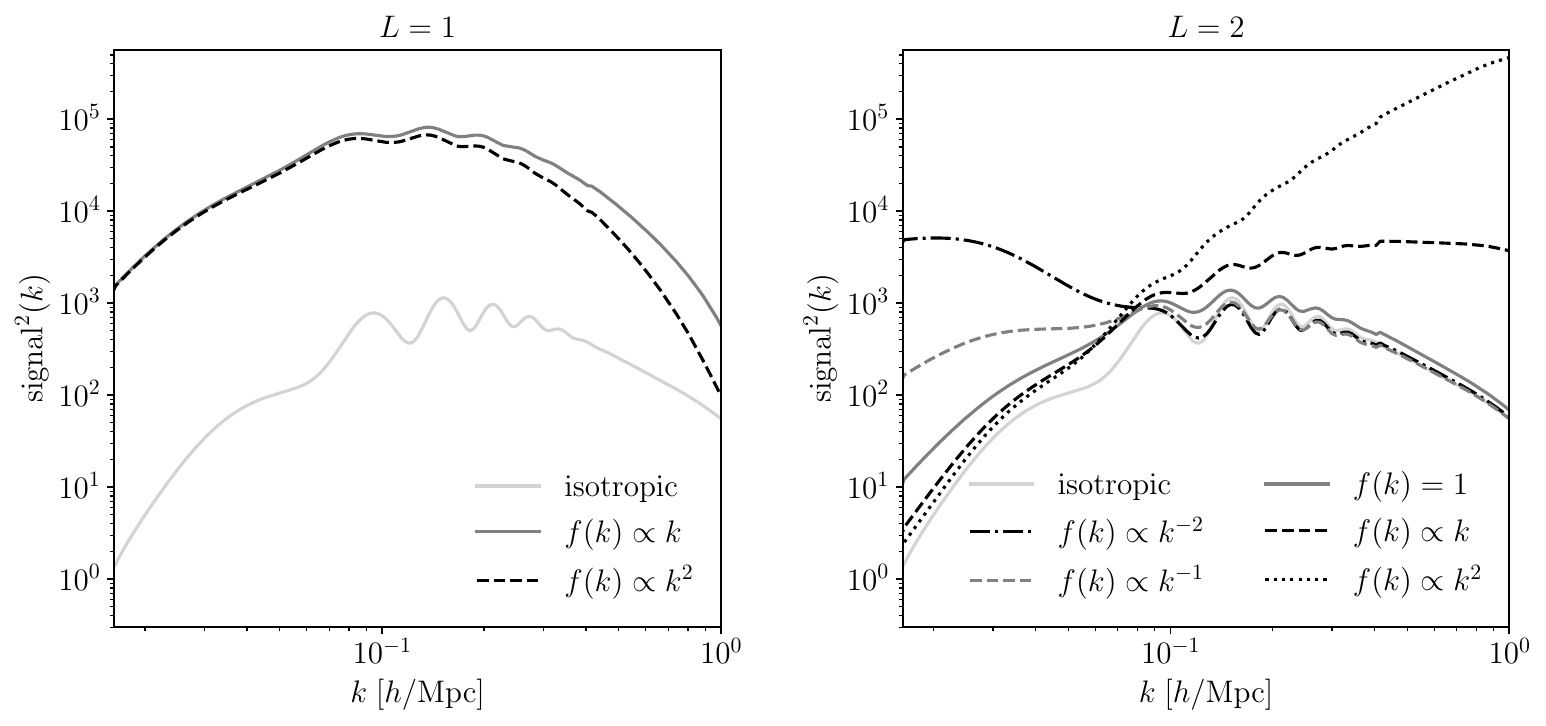}
     \caption{Signal $\left( P^{\rm A}(\vecks,\veckl) \right)^2$ integrated over angles and over long modes, for a correlation sourced by nonlinear biased clustering in the presence of an anisotropic modulation of the power spectrum, with $L=1$ (\emph{left panels}) and $L=2$ (\emph{right panels}). The amplitudes of the anisotropic contribution have been set to a larger value of 1, to visually appreciate the effect with respect to the purely isotropic case.}
     \label{fig:signal2_anisotropiesO1}
 \end{figure}

Figure~\ref{fig:signal2_anisotropiesO1} shows the signal squared, $\left( P^{\rm A}(\vecks,\veckl) \right)^2$ as a function of the short mode, both for the dipolar and the quadrupolar modulation.
The amplitude of the new term has been set to order $\mathcal{O}(1)$.

It can be seen how, for the dipolar modulation (and the modeling considered here), there is a scale dependent change, which modifies the amplitude especially at large scales. For the quadrupolar, on the other hand, the change varies depending on the power of the $f(k)$ considered, making different part of the analysis more sensitive to different models.

\subsubsection{Dipolar distortion $L=1$}

In~\cite{Planck2013,Planck2015a} the dipolar asymmetry was found to be $\sim 6-7$ \% on angular scales $\ell \lesssim 64$.
A dipolar distortion $L=1$ in the CMB can be described by a multiplicative modulation $T(\vhat{n})=T_{\rm iso}(\vhat{n})\left(1+A\vhat{n}\cdot\vhat{p}\right)$ with $\vhat{p}$ the preferred direction of the modulation, and $A \sim 0.07$~\cite{Gordon2005}. The power spectrum acquires the extra contribution
\begin{equation}
    \Qfenicio(\vec{k}) = \sum_M g_{1M}f(k)Y_{1M}(\vhat{k}) = A f(k) \frac{1}{2} \sqrt{\frac{3}{\pi}} \left( \cos\theta -\sqrt{2} \sin\theta \cos\phi \right) \, ,
\end{equation}
using $g_{1,-M}=(-1)^M g_{1M}^*$ and assuming for simplicity $g_{10}=g_{11}=A$. The following cases will be studied,
\begin{equation}
    f(k)=1-\frac{k}{k_{\rm A}} \, , \qquad \left(1-\frac{k}{k_{\rm A}}\right)^2 \, ,
\end{equation}
These parametrizations were introduced in~\cite{Shiraishi2016} as a heuristic model, to capture the main qualitative features emerging from CMB and quasar abundance observations, i.e.,~a large-scale dipolar asymmetry that rapidly decays and vanishes at $k \sim k_{\rm A} = 1 \ \Mpc^{-1}$.

\begin{figure}[h!]
     \centering
     \includegraphics[width=\textwidth]{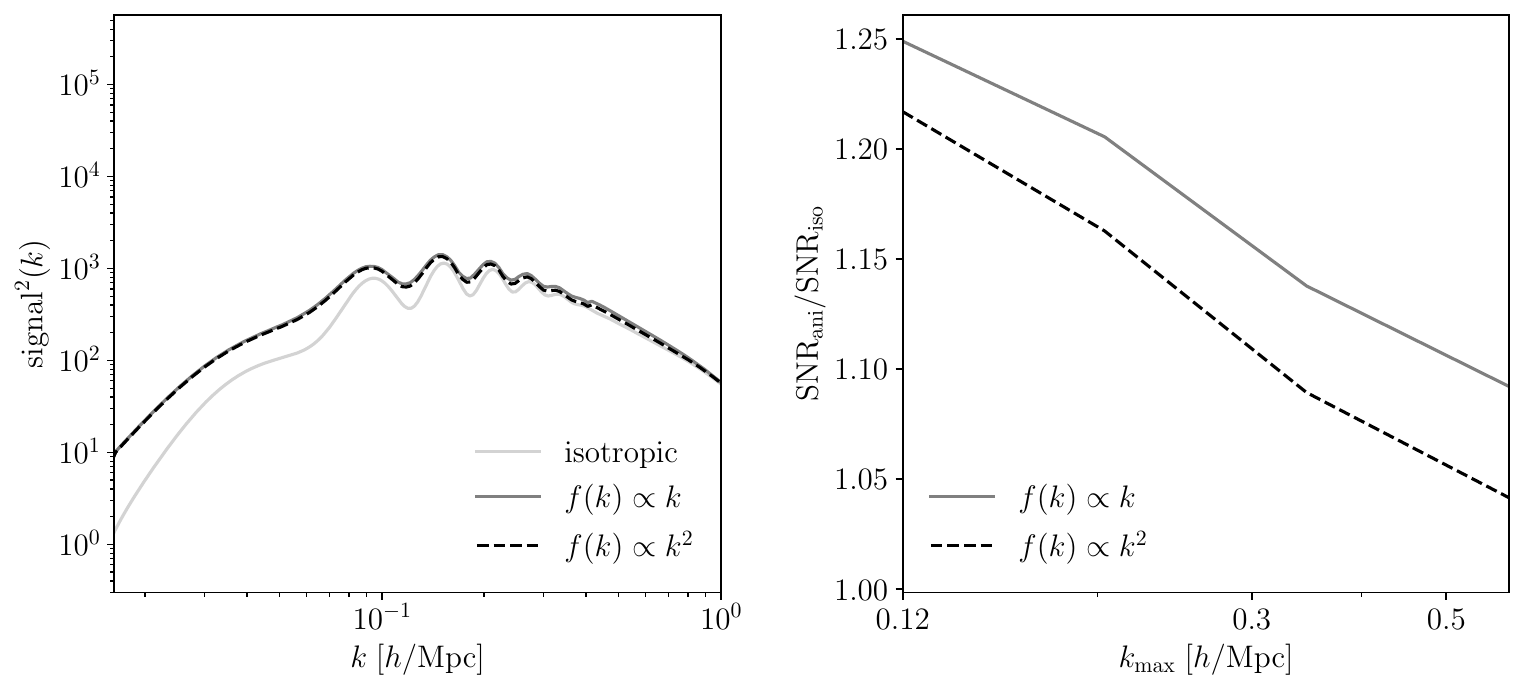}
     \caption{\emph{Left panel}. Signal $\left( P^{\rm A}(\vec{k},\vec{K}) \right)^2$ integrated over angles and over long modes, for a correlation sourced by nonlinear biased clustering in the presence of an anisotropic modulation of the power spectrum, with $L=1$. \emph{Right panel}. Ratio of the anisotropic SNR to the isotropic case. The amplitudes are the maximum ones allowed by Planck constraints, $A_{1M}=0.07$~\cite{Planck2015}. As before, the survey volume is $100 \ ({\rm Gpc}/h)^3$ centered at $z=1$, with tracers $n_{g,\tA}=10^{-3} \ (h/\Mpc)^3$, $n_{g,\tB}=2n_{g,\tA}$ and biases ($\bltA=3$, $\bltB=1.1$). The dipolar modulation results in a order few percent enhancement of the SNR, depending on the maximum scale $\ksmax$ used in the analysis.}
     \label{fig:plot_anisotropyL1}
 \end{figure}

Figure~\ref{fig:plot_anisotropyL1} shows the signal squared and the SNR in the presence of dipolar anisotropies, having set the amplitude of the anisotropic contribution to the maximum value currently allowed by Planck constraints, $A_{1M}=0.07$~\cite{Planck2015}.

As it can be seen, the effect is small. The right panel shows the ratio between the SNR for the anisotropic (dipolar) model and the isotropic signal from biased clustering.
It can be seen that the SNR gets enhanced with respect to the purely isotropic case. Depending on the significance of the isotropic case, there is in principle the opportunity to detect such a signal, and therefore be competitive with existing constraints from Planck. A more detailed analysis is needed, including a study on the trade-off between going to larger values of $\ksmax$ and having more general constraining power, but having a less relevant contribution from the dipolar modulation, which is most important on the largest scales. For this reason, the ratio to the isotropic SNR case approaches one on small scales, and there is not significant gain in pushing the analysis to very high $\ksmax$.

A possible direction to explore to detect anisotropic imprints within this framework can rely on a characterization of the scale dependence of the SNR, either as a function of the survey's short modes $k$ or as a function of $\ksmax$ -- provided nonlinear scales can be accurately modeled.

\subsubsection{Quadrupolar distortion $L=2$}

A quadrupolar distortion $L=2$ arises for example in models of inflation where the inflaton is coupled to a vector field with a nonvanishing vev~\cite{Dimastrogiovanni2010,Soda2012}. The term encoding the anisotropy is
\begin{equation}
    \Qfenicio(\vec{k}) = \sum_M g_{2M}f(k)Y_{2M}(\vhat{k}) = g_{2M} f(k) \frac{1}{4} \sqrt{\frac{5}{\pi}} \left[ 3\cos^2\theta-1 +\sqrt{6}\sin^2\theta\cos(2\phi) -\sqrt{6}\sin(2\theta)\cos\phi \right] \, ,
\end{equation}
with $g_{2,-M}=(-1)^M g^*_{2M}$~\cite{Planck2015} and for simplicity $g_{20}=g_{21}=g_{22}=g_{2M}$. Both the amplitudes and the function $f(k)$ are strongly dependent on the underlying inflationary model. In a model-agnostic approach, the following cases will be studied
\begin{equation}
    f(k)=1 \, , \qquad \left(\frac{k}{k_{\rm g}}\right)^{\pm 1} \, , \qquad \left(\frac{k}{k_{\rm g}}\right)^{\pm 2} \, ,
\end{equation}
with $k_{\rm g}=0.05 \ \Mpc^{-1}$ the pivot scale adopted in the Planck collaboration. Planck upper bounds give $|g_{2M}| \lesssim 10^{-2}$ for all these cases, except for $\left(k/k_{\rm g}\right)^{-2}$ for which the bound is stronger $g_{2,-2} \sim -10^{-4}$~\cite{Planck2015}.

\begin{figure}[h!]
     \centering
     \includegraphics[width=\textwidth]{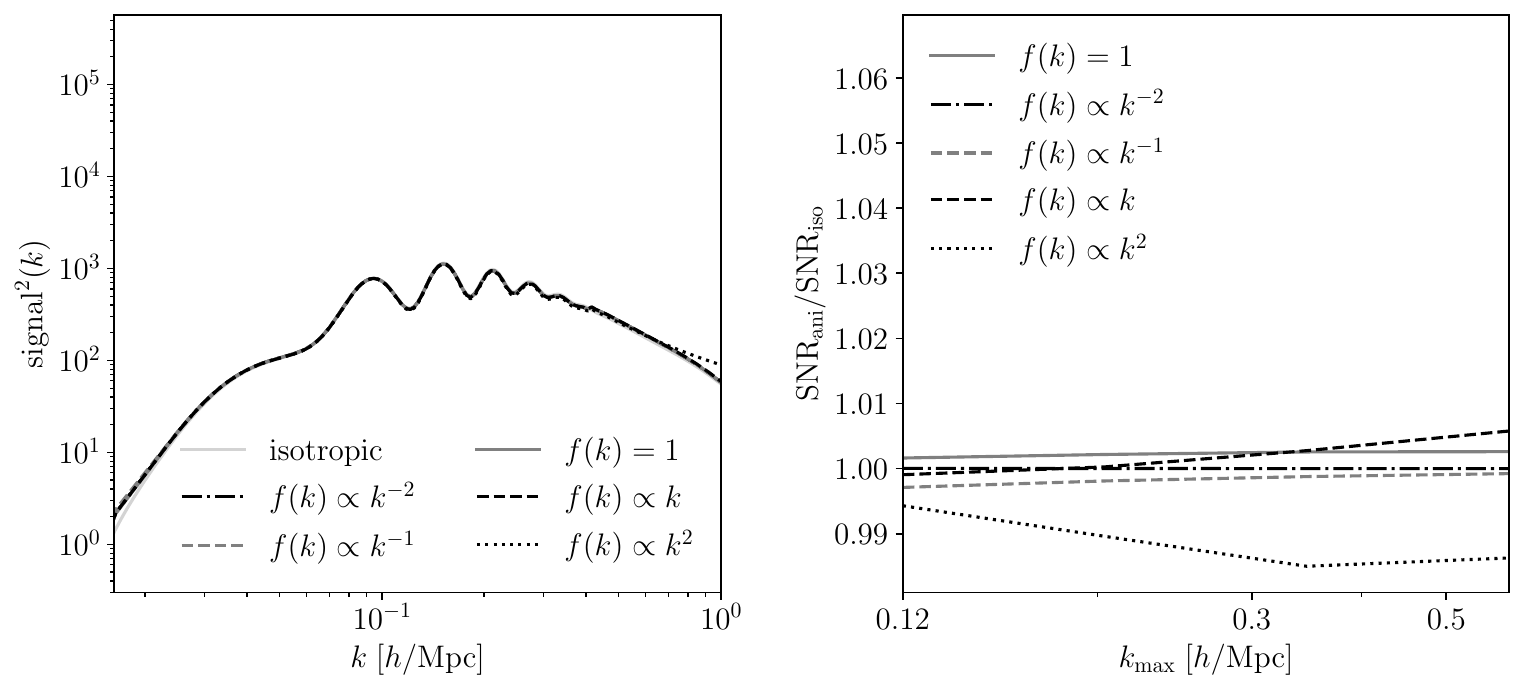}
     \caption{\emph{Left panel}. Signal $\left( P^{\rm A}(\vec{k},\vec{K}) \right)^2$ integrated over angles and over long modes, for a correlation sourced by nonlinear biased clustering in the presence of an anisotropic modulation of the power spectrum, with $L=2$. \emph{Right panel}. Ratio of the anisotropic SNR to the isotropic case. The amplitudes are the maximum ones allowed by Planck constraints, $g_{2,-2}=-10^{-4}$ and $g_{2M}=10^{-2}$ ($M=-1,0,1,2)$~\cite{Planck2015}. As before, the survey volume is $100 \ ({\rm Gpc}/h)^3$ centered at $z=1$, with tracers $n_{g,\tA}=10^{-3} \ (h/\Mpc)^3$, $n_{g,\tB}=2n_{g,\tA}$ and biases ($\bltA=3$, $\bltB=1.1$). The quadrupolar modulation essentially does not change the SNR with respect to the fully isotropic case, when within the current Planck limits.}
     \label{fig:plot_anisotropyL2}
 \end{figure}

Figure~\ref{fig:plot_anisotropyL2} shows the signal squared and the SNR in the presence of quadrupolar anisotropies, having set the amplitude of the anisotropic contribution to the maximum value allowed by Planck constraints~\cite{Planck2015}. As it can be seen, the modifications induced by such a level of quadrupolar anisotropy, on the scales of interest, are very small.

Therefore, it appears clear that Planck already constraints the amplitudes of the quadrupolar modulation to a level that will not realistically be reached by antisymmetric galaxy correlations. The right panel of the Figure shows how the SNR from this modulation is virtually unchanged.

\section{Conclusions}
\label{sec:conclusions}

This paper investigates one of the recently proposed observables for galaxy clustering, the antisymmetric galaxy cross-correlation.
The antisymmetric component of the two-point galaxy cross-correlation function arises when the small-scale power is modulated by a long wavelength field. Such a signal is sourced by the squeezed bispectrum of the two objects being correlated and the long mode, underlying field. This signal can be decomposed into longitudinal and transverse components, sourced by different physical mechanisms.

The first expression for this observable was given in~\cite{Dai2015}; this work provides a more accurate modeling, by adding redshift-space distortions, nonlinear gravitational evolution, second order bias expansion and primordial non-Gaussianity. Moreover, for the first time, the detectability of this signal is investigated from a quantitative point of view, by building a recipe for calculating the SNR and applying it to various examples.

In the standard $\Lambda$CDM scenario, an antisymmetric cross-correlation arises between different tracers of the underlying dark matter field, because of nonlinear biased clustering.

Beyond $\Lambda$CDM, the antisymmetric cross-correlation can pick up the signature of some exotic BSM scenarios.
In particular, this can happen for models where the two objects being correlated respond in a different way to the underlying field, and for models with anisotropic features inducing a privileged direction in the sky.

For these reasons, this observable can be a powerful tool to search for hints of new physics.
This work investigates the signature of some of these models.
A particularly interesting case is the imprint of vector modes~\cite{Dai2015} that arise, e.g.,~due to primordial magnetic fields~\cite{Widrow2002,Subramanian2015}, or in models of axion inflation~\cite{Freese1990,Pajer2013}. Other signatures of these fields are related to e.g.,~compensated isocurvature perturbations~\cite{Vanzan2023}.
Other examples are a two-component dark matter model as suggested in~\cite{Dai2015} and~\cite{Aguirre2004,Profumo2009,Adulpravitchai2011,Cyr-Racine2015,Herrero-Garcia2017,Herrero-Garcia2018}, and the extra signature coming from primordial anisotropies imprinted in the power spectrum after inflation~\cite{Shiraishi2016}.

These are just few examples, that are meant to demonstrate the potential of this new observable in testing both early and late Universe physics.
Its full potential for specific models for realistic surveys will be investigated in a future work.

\begin{acknowledgments}
We thank Nicola Bellomo, Colin Hill, Lam Hui, Donghui Jeong, Marc Kamionkowski, Ely Kovetz, Michele Liguori, Andrea Ravenni, Licia Verde for useful discussions.
AR acknowledges funding from Italian Ministry of University and Research (MUR) through the ``Dipartimenti di eccellenza'' project ``Science of the Universe''.
This work is supported in part by the MUR Departments of Excellence grant ``Quantum Frontiers''.
\end{acknowledgments}

\bibliography{bibliography}

\appendix

\section{Fourier space kernels}
\label{app:extra}

Keeping the full relation between redshift space and real space overdensity, Equations~\eqref{eq:RSD1}-\eqref{eq:RSD2}, including the Doppler term~\cite{Raccanelli2016} and selection effects, the first and second order redshift space kernels for biased tracers are, in Fourier space
\begin{equation}
    Z_1(\vec{k}) = b_1 +f\mu^2 -\frac{i\mu f \alpha}{kr} \, ,
\end{equation}
\begin{equation}
\begin{split}
    Z_2(\vec{p},\vec{q}) =& \frac{b_2}{2} +b_1F_2(\vec{p},\vec{q}) +b_{K^2}\left(\mu_{pq}^2-\frac{1}{3}\right) +f\mu^2 G_2(\vec{p},\vec{q}) \\
    & -\frac{i\mu f \alpha}{kr} G_2(\vec{p},\vec{q}) +\frac{k\mu f}{2}\left(\frac{\mu_p}{p}\left(b_1+f\mu_q^2\right) +\frac{\mu_q}{q}\left(b_1+f\mu_p^2\right)\right) \\
    & -\frac{if\alpha}{r} \left[ \frac{\mu_p}{p}\left(\frac{b_1}{2}+f\mu_q^2\right) +\frac{\mu_q}{q}\left(\frac{b_1}{2}+f\mu_p^2\right) \right] \\
    & -f^2\frac{\mu_p\mu_q}{pq r^2}\left( \alpha(\alpha-2)-\gamma \right) \, ,
\end{split}
\end{equation}
with $\vec{k}=\vec{p}+\vec{q}$, and
\begin{equation}
    \alpha \equiv \frac{r}{\Bar{n}}\frac{\partial\Bar{n}}{\partial r} +2 \, , \qquad\qquad \gamma \equiv \frac{r^2}{2\Bar{n}}\nabla^2\Bar{n} -3 \, .
\end{equation}

\section{Full expressions of the $\mathcal{F}$ functions}
\label{app:Fs}

These are the full expressions of the $\mathcal{F}(\vecks,\veckl)$ functions that appear in Equation~\eqref{eq:fullexprFK}. The superscripts refer to the corresponding order in the expansion in powers of the long mode over the short mode, $(\kl/\ks)$.
\begin{align}
    & \mathcal{F}_{\phi}^{(-1)}(\vecks,\veckl) = 42 T(\ks) f\mus^2 \left\{ \left[-2+\left(-4+\dlogPkdlogk\right) f \mus^2 \right] \musl +2 f \mus \mul \right\} \, , \\
    & \mathcal{F}_{\phi}^{(0)}(\vecks,\veckl) = -21 T(\ks) f \mus \left\{ -f \mus^3 \left[ -4 +\dlogPkdlogk +\left(24+\dPsecond \right) \musl^2 -9 \dlogPkdlogk \musl^2 \right] \right. \nonumber \\
    &\qquad \left. +4 \musl \mul -6 \left(-4+\dlogPkdlogk\right) f \mus^2 \musl \mul +2 \mus \left[ 1 +\left(-4+\dlogPkdlogk\right) \musl^2 -3 f \mul^2 \right] \right\} \, , \\
    & \mathcal{F}_{\phi}^{(1)}(\vecks,\veckl) = -6f\mus \left\{ 2\mul \left[ 1 +\left( -22+7\dlogPkdlogk-7\dTfirst \right) \musl^2 \right] \right. \nonumber \\
    &\qquad \left. +\mul \musl \left[ \dTfirst \left( -6+6\musl^2-7f\mul^2 \right) +\dlogPkdlogk \left(6-6\musl^2+7f\mul^2\right) -4\left(3-3\musl^2+7f\mul^2\right) \right] \right\} \nonumber \\
    &\qquad +7f T(\ks) \left\{ -6\musl\mul^2 +6\mus\mul \left[-2-2\left(-4+\dlogPkdlogk\right) \musl^2 +f \mul^2 \right] \right. \nonumber \\
    &\qquad \left. +9f\mus^3\mul \left[ -4+\dlogPkdlogk-9\dlogPkdlogk\musl^2+\left(24+\dPsecond\right) \musl^2 \right] \right. \nonumber \\
    &\qquad +\left. f\mus^4 \musl \left[ 72-192\musl^2+\dPthird\musl^2+\dPsecond(3-15\musl^2) +3\dlogPkdlogk (-9+29\musl^2) \right] \right. \nonumber \\
    &\qquad \left. -3 \mus^2 \musl \left[ -12+\left(24+\dPsecond\right) \musl^2 +24 f \mul^2 +\dlogPkdlogk (3-9\musl^2-6f\mul^2) \right] \right\} \, ,
\end{align}

\begin{align}
    & \mathcal{F}_{\phi\delta}^{(-1)}(\vecks,\veckl) = 42 T(\ks) f\mus \left[ \left(-2+\dlogPkdlogk\right) \mus \musl +2 \mul \right] \, , \\
    & \mathcal{F}_{\phi\delta}^{(0)}(\vecks,\veckl) = 21 T(\ks) f \left\{ \mus^2 \left[ -2+\dlogPkdlogk+\left(8+\dPsecond\right) \musl^2-5 \dlogPkdlogk \musl^2 \right] \right. \nonumber \\
    &\qquad \left. +4\mus\mul\musl \left(-2+\dlogPkdlogk\right) +2 \mul^2 \right\} \, , \\
    & \mathcal{F}_{\phi\delta}^{(1)}(\vecks,\veckl) = 7f \left\{ 6\mus \left[2 \mul-\left(4-\dlogPkdlogk+\dTfirst\right) \mus \musl \right] \right. \nonumber \\
    &\qquad \left. +T(\ks) \left[ \mus^2 \musl \left( 24-48\musl^2+\dPthird \musl^2+\dPsecond (3-9\musl^2) +3 \dlogPkdlogk (-5+11\musl^2) \right) \right. \right. \nonumber \\
    &\qquad \left. \left. +6 \mus \left( -2+\dlogPkdlogk-5 \dlogPkdlogk \musl^2+\left(8+\dPsecond\right) \musl^2 \right) \mul +6 \left(-2+\dlogPkdlogk\right) \musl \mul^2 \right] \right\} \, ,
\end{align}

\begin{align}
    & \mathcal{F}_{1,\phi}^{(-1)}(\vecks,\veckl) = -42 T(\ks) \musl \left[ -2+\left(-2+\dlogPkdlogk\right) f\mus^2 \right] \, , \\
    & \mathcal{F}_{1,\phi}^{(0)}(\vecks,\veckl) = 21 T(\ks) \left\{ 2+2\musl^2\left(-2+\dlogPkdlogk\right) \right. \nonumber \\
    &\qquad \left. +f \mus \left[ \mus \left(2-\dlogPkdlogk-\left(8+\dPsecond\right) \musl^2 +5 \dlogPkdlogk \musl^2\right) \right. \right. \nonumber \\
    &\qquad \left. \left. +2 \musl \mul \left(2-\dlogPkdlogk\right) \right] \right\} \, , \\
    & \mathcal{F}_{1,\phi}^{(1)}(\vecks,\veckl) = -6\musl \left[26 +28f\mus^2 +16\musl^2 -\dlogPkdlogk (10 +7f\mus^2 +4\musl^2) +\dTfirst (10 +7f\mus^2 +4\musl^2) \right] \nonumber \\
    &\qquad -7 T(\ks) \left\{ 3\musl \left[6 + \left(8+\dPsecond\right) f\mus^2 -\dlogPkdlogk (3+5f\mus^2) \right] \right. \nonumber \\
    &\qquad \left. +\musl^3 \left(-24 -48f\mus^2 +\dPthird f \mus^2 -3\dPsecond (1+3f\mus^2) +3 \dlogPkdlogk (5+11f\mus^2)\right) \right. \nonumber \\
    &\qquad \left. +3f\mus\mul \left(-2+\dlogPkdlogk\right) +3f\mus\mul\musl^2 \left(8-5\dlogPkdlogk+\dPsecond\right) \right\} \, ,
\end{align}

\begin{align}
    & \mathcal{F}_{1,\phi\delta}^{(-1)}(\vecks,\veckl) = -42 T(\ks) \dlogPkdlogk \musl \, , \\
    & \mathcal{F}_{1,\phi\delta}^{(0)}(\vecks,\veckl) = -7 T(\ks) \left(3 \dPsecond \musl^2 -3 \dlogPkdlogk (-1+\musl^2) \right) \, , \\
    & \mathcal{F}_{1,\phi\delta}^{(1)}(\vecks,\veckl) = 7 \musl \left[ 6 \left(2-\dlogPkdlogk+\dTfirst\right) \right. \nonumber \\
    &\qquad \left. -T(\ks) \left( 3 \dPsecond -3 \dPsecond \musl^2 +\dPthird \musl^2 +3 \dlogPkdlogk (-1+\musl^2) \right) \right] \, ,
\end{align}

\begin{align}
    & \mathcal{F}_{\phi,2}^{(1)}(\vecks,\veckl) = -42 \musl \left(-2+\dlogPkdlogk-\dTfirst\right) \, , \\
    & \mathcal{F}_{\phi,K^2}^{(1)}(\vecks,\veckl) = -28 \musl \left( 8+\dTfirst-12\musl^2-3\dTfirst\musl^2+\dlogPkdlogk(-1+3\musl^2)\right) \, .
\end{align}

\section{Bias parameters}
\label{app:biases}

In this paper, the second order biases have been linked to the linear order one by means of the fitting formula and the LIMD relation respectively~\cite{Lazeyras2015,Desjacques2016}
\begin{align}
    & b_2(b_1) = 0.412 -2.143 b1 +0.929 b1^2 +0.008 b1^3 \, , \\
    & b_{K^2}(b_1) = -\frac{2}{7}(b_1-1) \, ,
\end{align}
and the universal mass function relations
\begin{alignat}{2}
    & b_{\phi} &&= 2\delta_{\rm cr}(b_1-1) \, , \\
    & b_{\phi\delta} &&= -b_{1,L}+\delta_{\rm cr}b_{2,L} +b_{\phi} \nonumber \\
    & &&= 1-b_1+b_{\phi}+\left(\frac{8}{21}(1-b_1)+b_2\right)\delta_{\rm cr} \, ,
\end{alignat}
with $\delta_{\rm cr}=1.686$, and the Lagrangian biases being connected to the Eulerian ones as~\cite{Mo1995,Mo1996,Desjacques2018}
\begin{align}
    & b_{1,L}=b_1-1 \, , \\
    & b_{2,L} = b_2 -\frac{8}{21} \left( b_1-1 \right) \, .
\end{align}

For generic type PnG, one needs to take a step back to the definition of the bias parameters as encoding the response of the galaxy overdensity to a change in the initial conditions, e.g.,~if one parametrizes this change as a rescaling of the initial density perturbation by $\left( 1+2\epsilon k^{-\alpha} \right)$, then
\begin{equation}
    b^L_{\psi\delta^N} = \frac{1}{N!} \frac{1}{\bar{n}_g} \left.\frac{\partial^{N+1} \bar{n}_g}{\partial\Delta^N \partial\epsilon}\right|_{\Delta=0,\epsilon=0} \, ,
\end{equation}
where $\Delta$ is the long wavelength component of the overdensity field~\cite{Desjacques2011,Schmidt2012,Desjacques2016}.
The Eulerian bias parameters $b_{\psi}$ and $b_{\psi\delta}$ are obtained as~\cite{Giannantonio2009,Karagiannis2018}
\begin{equation}
    b_{\psi} = \left[ b_{\phi} +4\left( \frac{\partial \ln\sigma^2_{-\alpha/2}}{\partial \ln\sigma^2} -1 \right) \right] \frac{\sigma^2_{-\alpha/2}}{\sigma^2} \, ,
\end{equation}
and
\begin{equation}
    b_{\psi\delta} = A \left[ \delta_{\rm cr}b_2 +\frac{17}{21} 2\delta_{\rm cr}(b_1-1) +(b_1+1)\left( 2\frac{\partial \ln\sigma^2_{-\alpha/2}}{\partial \ln\sigma^2} -3 \right) +2 \right] \frac{\sigma^2_{-\alpha/2}}{\sigma^2} \, ,
\end{equation}
with $\alpha=0$, $A=1$ for the local case and $\alpha=1$, $A=-3$ for the orthogonal case, and
\begin{equation}
    \sigma^2_n = \int\frac{d^3\vec{k}}{(2\pi)^3} k^n P(k) \left| W(k) \right|^2 \, .
\end{equation}

\section{Estimator}
\label{app:estimator}

An estimator for the amplitude of the long mode was first proposed in~\cite{Jeong2012} and then applied in~\cite{Dai2015} to the antisymmetric case, in the null hypothesis where $\langle\delta_1(\vec{k}_1)\delta_2(\vec{k}_2)\rangle$ is vanishing in the absence of the modulating field. In the case of biased clustering, however, the situation is different, and the procedure outlined there needs to be generalized.

To single out the antisymmetric part of the signal, the quantity of interest is, in discretized form~\cite{Zhou2020}
\begin{equation}
    \frac{1}{2} \left[ \delta_1(\vec{k}_1)\delta_2(\vec{k}_2)-\delta_1(\vec{k}_2)\delta_2(\vec{k}_1) \right] = V_s \delta_{\vec{k}_1,\vec{k}_2,\vec{K}} \delta^*(\vec{K}) f_L^{\rm A}(\vec{k}_1,\vec{k}_2) \vhat{K}\cdot\left(\vec{k}_1-\vec{k}_2\right) \, ,
\end{equation}
where $V_s \delta_{\vec{k}_1,\vec{k}_2,\vec{K}} \xrightarrow{V \xrightarrow{} \infty} (2\pi)^3 \delta^{(3)}(\vec{k}_1+\vec{k}_2+\vec{K})$.

Each pair $\vec{k}_1$, $\vec{k}_2$ provides an estimator (changing sign of $\Vec{K}$)
\begin{equation}
    \widehat{\delta(\vec{K})} = \frac{1}{2} \left[ \delta_1(\vec{k}_1)\delta_2(\vec{k}_2)-\delta_1(\vec{k}_2)\delta_2(\vec{k}_1) \right] \left[ f_L^{\rm A}(\vec{k}_1,\vec{k}_2) \vhat{K}\cdot\left(\vec{k}_1-\vec{k}_2\right) \right]^{-1} \, .
\end{equation}
The variance of the antisymmetrized combination of density fluctuations is
\begin{equation}
\begin{split}
    & \frac{1}{4}\langle \left(\delta_1(\vec{k}_1)\delta_2(\vec{k}_2)-\delta_1(\vec{k}_2)\delta_2(\vec{k}_1)\right) \left(\delta_1(\vec{k}_1\pr)\delta_2(\vec{k}_2\pr)-\delta_1(\vec{k}_2\pr)\delta_2(\vec{k}_1\pr)\right) \rangle + \\
    & \qquad\qquad -\langle \delta_1(\vec{k}_1)\delta_2(\vec{k}_2)-\delta_1(\vec{k}_2)\delta_2(\vec{k}_1) \rangle \langle \delta_1(\vec{k}_1\pr)\delta_2(\vec{k}_2\pr)-\delta_1(\vec{k}_2\pr)\delta_2(\vec{k}_1\pr) \rangle \\
    =& \frac{1}{4} \left[ P_1(k_1)P_2(k_2)+P_1(k_2)P_2(k_1)-2P_{12}(k_1)P_{12}(k_2) \right] \left( \delta^D_{\vec{k}_1+\vec{k}_1\pr}\delta^D_{\vec{k}_1+\vec{k}_2\pr} -\delta^D_{\vec{k}_1+\vec{k}_2\pr}\delta^D_{\vec{k}_1\pr+\vec{k}_2} \right) \, ,
\end{split}
\end{equation}
where $\vec{k}_2=\vec{K}-\vec{k}_1$ and $\vec{k}_2\pr=\vec{K}-\vec{k}_1\pr$. The term $\left[ f_L^{\rm A}(\vec{k}_1\pr,\vec{k}_2\pr) \vhat{K}\cdot\left(\vec{k}_1\pr-\vec{k}_2\pr\right) \right]^{-1}$ picks up a negative sign from the second combination of Dirac deltas, so that the overall variance gets a factor 2 and becomes
\begin{equation}
    \frac{V_s}{2} \left[ f_L^{\rm A}(\vec{k}_1,\vec{k}_2) \vhat{K}\cdot\left(\vec{k}_1-\vec{k}_2\right) \right]^{-2} \left( P_1(k_1)P_2(k_2)+P_1(k_2)P_2(k_1)-2P_{12}(k_1)P_{12}(k_2) \right) \, .
\end{equation}
The optimal estimator is obtained by summing over all modes with inverse variance weightings
\begin{equation}
\begin{split}
    & \widehat{\delta(\vec{K})} = P_n(K) \sum_{\vec{k}} \frac{ \left[ f_L^{\rm A}(\vec{k}_1,\vec{k}_2) \vhat{K}\cdot\left(\vec{k}_1-\vec{k}_2\right) \right] }{ \frac{V_s}{2} \left( P_1(k_1)P_2(k_2)+P_1(k_2)P_2(k_1)-2P_{12}(k_1)P_{12}(k_2) \right) } \frac{1}{2} \left[ \delta_1(\vec{k}_1)\delta_2(\vec{k}_2)-\delta_1(\vec{k}_2)\delta_2(\vec{k}_1) \right] \, , \\
    & P_n(K) = \left[ \sum_{\vec{k}} \frac{ \left[ f_L^{\rm A}(\vec{k}_1,\vec{k}_2) \vhat{K}\cdot\left(\vec{k}_1-\vec{k}_2\right) \right]^2 }{ \frac{V_s}{2} \left( P_1(k_1)P_2(k_2)+P_1(k_2)P_2(k_1)-2P_{12}(k_1)P_{12}(k_2) \right) } \right]^{-1} \, .
\end{split}
\end{equation}

Since $\langle\left|\widehat{\delta(\vec{K})}\right|^2\rangle=V_s\left(P(K)+P_n(K)\right)$, if one parametrizes $P(K)=AP_f(K)$ with a fiducial power-spectrum $P_f(K)$, each $\vec{K}$ provides an estimator for the amplitude
\begin{equation}
    \Hat{A}_{\vec{K}} = P_f(K)^{-1} \left( V_s^{-1} \left|\widehat{\delta(\vec{K})}\right|^2-P_n(K) \right) \, ,
\end{equation}
so the optimal estimator is
\begin{equation}
    \Hat{A} = \sigma^2 \sum_{\vec{K}} \frac{P_f(K)}{2\left(P_n(K)\right)^2} \left( V_s^{-1}\left|\widehat{\delta(\vec{K})}\right|^2-P_n(K) \right) \, ,
\end{equation}
\begin{equation}
    \sigma^{-2} = \sum_{\vec{K}} \frac{\left(P_f(K)\right)^2}{2\left(P_n(K)\right)^2} \, .
\end{equation}

The quantity $\sigma$ roughly represents the sensitivity to the amplitude of the underlying long mode power spectrum, and therefore can be used to get an estimate of the detection threshold of the modulation effect.

The sum over Fourier modes becomes $\sum_{\vec{k}} \longmapsto V/(2\pi)^3 \int d^3\vec{k}$.
For the short modes, $V=V_s$ the survey volume. In order to restrict to squeezed configurations only, at a given long mode $\kl$, the lower integration limit is set to $\ksmin = 10K$, where $10$ is the (arbitrarily chosen) minimum squeezing factor.
As for the long modes, the sum over modes should account for all the large scales that in principle can modulate the two-point function on the scales $\vec{k}_1$, $\vec{k}_2$. Modes that are much larger than the survey will be degenerate with the background, therefore $V \simeq V_s$ as well, and the lower integration limit can be taken to be of order the fundamental wavevector of the survey, $\klmin \simeq 2\pi / V_s^{1/3}$.
Both integrations run up to $k_{\rm max}$. In practice, given the squeezing requirement, the integration over long modes stops at $\klmax = k_{\rm max}/10$.

\end{document}